\documentclass[aps,floats,showpacs,nofootinbib, twocolumn]{revtex4-1} 

\usepackage{amssymb}
\usepackage{amsmath}

\usepackage{sans}

\usepackage{graphicx}

\usepackage{graphics}
\usepackage{dcolumn}
\usepackage{color}
\usepackage{rotate}
\usepackage{fancyhdr}
\usepackage{hyperref}
\usepackage{indentfirst}

\usepackage{umoline}
\usepackage{ulem}

\newcommand{\apjl}{Astrophys.\ J.\ Lett.\  }

\usepackage{physics}
\usepackage{booktabs}
\usepackage{fontawesome}

\newcommand{\beq}{\begin{equation}}
\newcommand{\eeq}{\end{equation}}
\newcommand{\be}{\begin{eqnarray}}
\newcommand{\ee}{\end{eqnarray}}
\newcommand{\ba}{\begin{eqnarray}}
\newcommand{\ea}{\end{eqnarray}}
\newcommand{\no}{\nonumber}

\newcommand{\iu}{\mathrm{i}}
\newcommand{\e}{\mathrm{e}}

\definecolor{darkred}{rgb}{.743,0,0}

\begin{document}
\title{Host group degeneracy in gravitational lensing time delay determination of $H_0$
}

\author{Luca Teodori}\email{luca.teodori@weizmann.ac.il}
\affiliation{Weizmann Institute, Department of Particle Physics and Astrophysics, Rehovot, Israel 7610001}

\author{Kfir Blum}\email{kfir.blum@weizmann.ac.il}
\affiliation{Weizmann Institute, Department of Particle Physics and Astrophysics, Rehovot, Israel 7610001}

\begin{abstract}
Massive elliptical galaxies, that serve as lenses in gravitational lensing time delay measurements of the Hubble parameter $H_0$, often reside in a host group.  
We consider degeneracies in the modeling of the group halo. When the group effect on imaging can be summarized by its flexion (the next order term beyond shear in the tidal expansion), the posterior likelihood map can develop disjoint local minima, associated with an approximate discrete symmetry of a dominant flexion term. Monte-Carlo Markov Chain (MCMC) algorithms that are not designed to explore a rich posterior landscape can miss some of the minima, introducing systematic bias. We study mock data and demonstrate that the bias in $H_0$ can exceed 10\%, and pulls the inference value of $H_0$ above its truth value, for a reason that can be traced to the structure of a mismodeled flexion term. MCMC algorithms that are designed to cope with a rich posterior landscape can uncover the structure. If the group is X-ray bright enough, X-ray data may also help to resolve the degeneracy, by pinpointing the group's center of mass. Finally, we show that some implementations in the literature used an inaccurate kinematical prior, mis-modeling the group velocity dispersion by as much as $20\%$.
\end{abstract}

\maketitle
\tableofcontents
\section{Introduction}
Gravitationally lensed quasars allow a determination of the Hubble parameter $H_0$~\cite{10.1093/mnras/128.4.307,Suyu:2012aa,Treu:2016ljm,Suyu:2016qxx,Grillo:2020yvj}, and the results of such  measurements~\cite{Rusu:2019xrq,Birrer:2018vtm,DES:2019fny,Chen:2019ejq, Wong:2019kwg,Millon:2019slk} were widely considered as tests of the cosmological  model~\cite{SHOES,Verde:2019ivm,DiValentino:2021izs}. 
However, systematic degeneracies are a limiting factor in the interpretation of lensing data~\cite{Falco1985,1991ApJ...373..354K,Kochanek:2002rk,Liesenborgs:2012pu,Schneider_2013,Kochanek:2020crs,Fleury:2021tke,Teodori:2022ltt}.  
Relaxing some of the modeling assumptions made in~\cite{Rusu:2019xrq,Birrer:2018vtm,DES:2019fny,Chen:2019ejq, Wong:2019kwg,Millon:2019slk}, a possible tension between the value of $H_0$ inferred via lensing and via cosmic microwave background (CMB) and large-scale structure (LSS)~\cite{Akrami:2018vks,DAmico:2019fhj,Ivanov:2019pdj} analyses may be attributed to a core feature in the density profile in or around the lenses~\cite{Schneider_2013,Blum:2020mgu,Birrer:2020tax,Blum:2021oxj}. 

\begin{figure}
    \centering
    \includegraphics[scale=0.4]{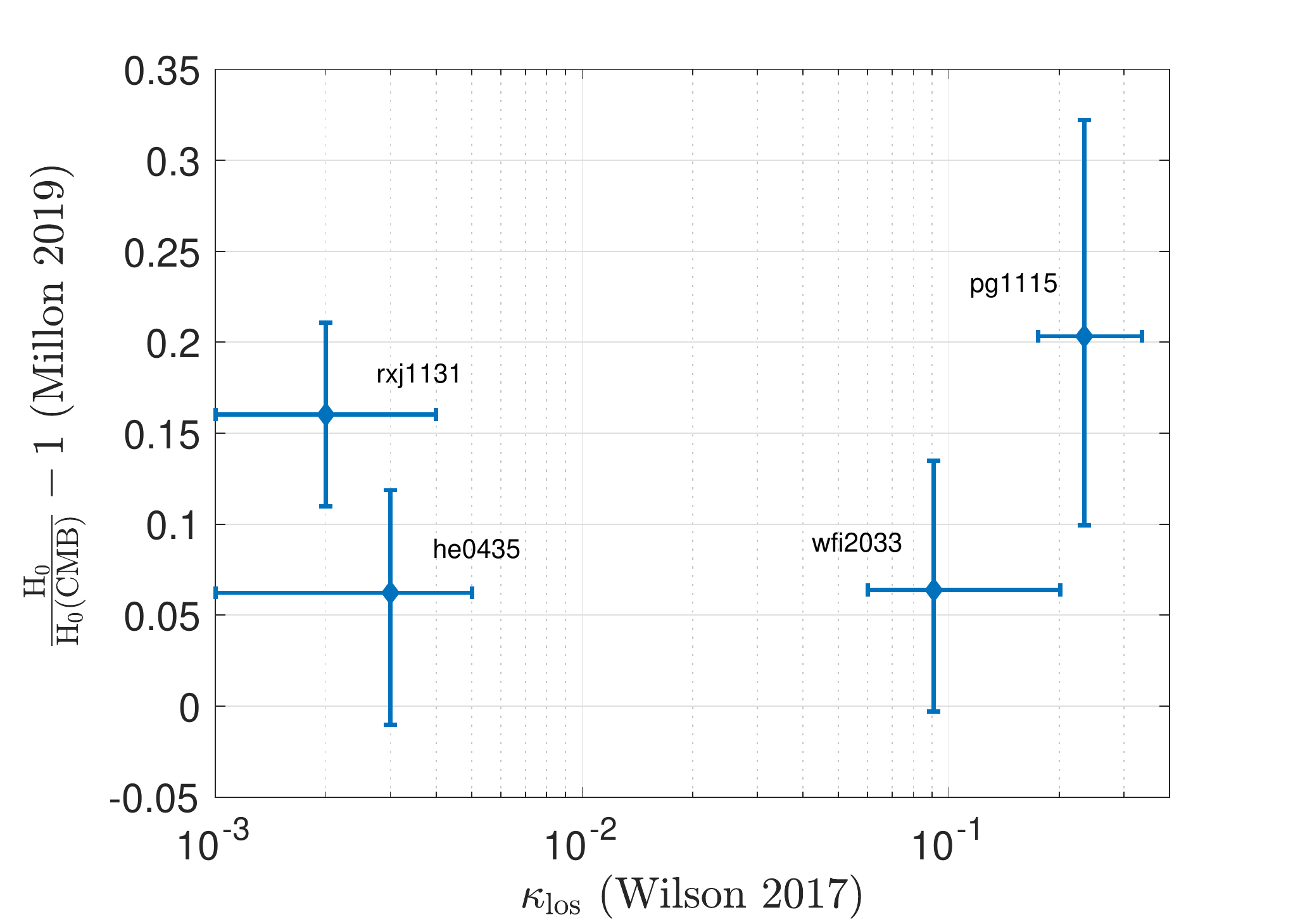}
    \caption{Lensing-inferred $H_0$ (from~\cite{Millon:2019slk}, with ``$H_0$(CMB)" taken from~\cite{Akrami:2018vks}) vs. estimated LOS convergence (from~\cite{Wilson:2017apg}).}
    \label{fig:H0k}
\end{figure}

A core feature is an approximate mass-sheet degeneracy (MSD)~\cite{Falco1985}. It could be an intrinsic characteristic of the lens galaxy itself, on distances of dozens of kpc~\cite{Blum:2020mgu}. 
However, the effect could also come from larger scales. On intermediate scales, in between cosmology and lens internal structure, it is noteworthy that massive galaxies like the lenses of~\cite{Rusu:2019xrq,Birrer:2018vtm,DES:2019fny,Chen:2019ejq, Wong:2019kwg,Millon:2019slk} are often members of a group\footnote{Ref.~\cite{Wilson:2016hcs} finds that half of their sample of 26 galaxy lenses can be associated with a group.}, and the dark matter halo of a host group may act to some extent as a core-MSD. 
Ref.~\cite{Wilson:2016hcs,Wilson:2017apg} studied the impact of host and line of sight (LOS) groups on lensing systems. Among the systems considered, PG1115+080, RXJ1131-1231, HE0435-1223, and WFI2033-4723 featured in the $H_0$ campaign of~\cite{Millon:2019slk}\footnote{At least two additional systems from~\cite{Millon:2019slk}  (SDSS1206~\cite{Birrer:2018vtm} and DES J0408~\cite{DES:2020ohe}), do not feature in~\cite{Wilson:2016hcs,Wilson:2017apg}, but are known to also be associated with host groups. Thus at least 6 out of the 7 lensed quasars from~\cite{Millon:2019slk} involve a group.}. We show these systems in Fig.~\ref{fig:H0k}. 
Interestingly, PG1115+080 yields a high central value of $H_0$ ($81^{+8}_{-7}$~km/s/Mpc)~\cite{Millon:2019slk}; at the same time, this lens resides in a massive group, inducing convergence $\kappa_{\rm g}\approx0.2$~\cite{Wilson:2017apg}. The group convergence goes directly into the inference of $H_0$, via $\delta H_0/H^{\rm truth}_0\approx-\delta\kappa_{\rm g}$, where $\delta\kappa_{\rm g}=\kappa_{\rm g}^{\rm truth}-\kappa_{\rm g}^{\rm model}$ is any error in the model determination of $\kappa_{\rm g}$, and $\delta H_0=H_0^{\rm truth}-H_0^{\rm model}$. We do not know if the lensing results for PG1115+080 are indeed biased by its group modeling (and will not make such a claim in this paper); but clearly, it is important to understand to what accuracy can lensing analyses determine $\kappa_{\rm g}$. 

In this paper we explore the group effect. The outline and main results are as follows. 
In Sec.~\ref{s:group} we consider simple analytic estimates. We show that when the distance separating the group's centroid from the primary lens is large in comparison to the primary lens's Einstein angle, attempts to model the group halo effects based on imaging data suffer from an approximate MSD, where the mass sheet comes from the group halo itself. This version of the MSD persists even when second-order tidal effects (flexion terms~\cite{Goldberg:2004hh,Bacon:2005qr}) are clearly detectable. Kinematics and astrometry of tracer galaxies are needed to break the degeneracy. 

In Sec.~\ref{s:mock} we perform numerical mock data analysis, motivated by a realistic example. We show that the posterior likelihood of the lensing reconstruction problem exhibits three disjoint minima. These minima reflect an approximate discrete degeneracy, related to the transformation properties of a dominant flexion term under coordinate rotations around the primary lens. We show that a naive implementation of a commonly used Monte-Carlo Markov Chain (MCMC) algorithm tends to fall into one of the minima, missing the others. Interestingly, and dangerously, the best fit $H_0$ obtained in any one of the wrong minima is systematically biased high, for a reason that we explain in Sec.~\ref{s:Bias}. The size of the $H_0$ bias can reach $\sim10\%$. An MCMC implementation that is especially designed to probe a rich likelihood landscape, uncovers the full degeneracy structure. 

We summarize in Sec.~\ref{s:sum}. 
Many details are kept to appendices, including relevant  calculations that exist in the literature (for completeness of some of our arguments), and also a few results that we did not see elsewhere. In App.~\ref{app:flexion} we spell out some details of the lensing potential of a Navarro-Frenk-White (NFW) profile~\cite{Navarro:1996gj}. 
In App.~\ref{app:kin} we review the relation between group member velocity dispersion and halo model. We note that some lensing studies implemented an incorrect kinematics prior for the NFW group model. Depending on selection cuts and some other factors, the systematic error in the interpretation of group velocity dispersion can reach $\sim20\%$. 
In App.~\ref{ss:probfalse} we give a rough estimate of the probability that an MCMC will actually fall into a wrong minimum; for the example of PG1115+080~\cite{Millon:2019slk} this probability is not large, on the order of $\sim10\%$. 
In App.~\ref{app:d} we attempt a (very) crude estimate of the fraction of strong galaxy lensing systems where flexion degeneracy may be expected to become a concern, finding this fraction to be in the ballpark of $\sim10\%-30\%$.
In App.~\ref{app:1} we collect expanded versions of MCMC triangle plots along with some sanity checks of our analysis.

We should comment that throughout the analysis, we do not consider stellar kinematic measurements of the primary lens itself (see~\cite{Yildirim:2021wdd,Shajib:2023uig} for state-of-the-art). Primary lens kinematics can constrain the group-induced MSD, if it can reach a sensitivity for over-all scaling of the primary lens mass model at the level of $1-\delta\kappa_g$.

\section{Host group as a core-MSD}\label{s:group}
In this section we provide simple analytic estimates that clarify the group effect on the lensing problem. The discussion is useful in understanding features of the numerical analysis of the next section. 

In systems like those considered in~\cite{Millon:2019slk}, the Einstein angle of the primary lens is of the order of $\theta_{\rm E}\sim1''$, and most of the imaging information lies at angular separation $\theta\sim\theta_{\rm E}$ around the lens. This angular scale projects onto a physical separation of the order of $\lesssim10$~kpc at the redshifts $z_{\rm l}\sim0.1-1$ of typical lenses. In comparison, a typical separation of any galaxy (with the possible exception of the brightest group galaxy (BGG)) from the host group's center of mass is $\gtrsim100$~kpc, that is, angular distance $h\gtrsim10''$. Therefore, for a rough estimate of the impact of a group on imaging analyses, it is sensible to expand the group's lensing potential in powers of $\theta/h$.

We adopt complex notation for 2D angle vectors on the sky~\cite{Schramm:1994he,Schneider:2007ks}, defining, e.g., 
\be\theta&=&\theta_1+\iu\theta_2,\;\;\;\;\beta\;=\;\beta_1+\iu\beta_2,\ee
etc. We set the origin of coordinates at the center of the primary lens. With this formalism, the lensing equation can be written as
\be\label{eq:lenseq}\beta&=&\theta-\alpha_{\rm l}(\theta)-\Delta\alpha(\theta)-\kappa_{\rm ext}\theta-\gamma_{\rm ext}\theta^*.\ee
Here, $\alpha_{\rm l}(\theta)$ is the deflection angle due to the primary lens, $\Delta\alpha(\theta)$ is the deflection due to the host group or cluster, and $\beta$ is the source position. The external convergence and shear, $\kappa_{\rm ext}$ and $\gamma_{\rm ext}$, contain a combination of different line of sight (LOS) contributions\footnote{For example, in terms of observer-source, observer-lens, and lens-source LOS terms, we have $\kappa_{\rm ext}=\kappa_{\rm s}+\kappa_{\rm l}-\kappa_{\rm ls}$. We also absorb some external convergence and shear terms into the definition of the ``primary lens", ``group", and ``source position" terms~\cite{Teodori:2022ltt}.}.

In what follows, when we refer to a host group, we consider the group's central dark matter halo, rather than individual member galaxies. For a group located at center of mass position $\e^{\iu\phi}h$, with $h\gg|\theta|$, $\Delta\alpha$ can be expanded as a power series in $\theta$~\cite{Schneider:2007ks} (see App.~\ref{app:flexion} for more details):
\be\label{eq:aexp}\Delta\alpha(\theta)&=&\Delta\beta +\Delta\kappa\,\theta+\Delta\gamma\,\theta^* \no\\
&+&\frac{1}{4}F^*\theta^2+\frac{1}{2}F\theta\theta^*+\frac{1}{4}G\theta^{*2}+\ldots\,.\ee
In the case of an axisymmetric group halo profile, it is possible to decompose the expansion coefficients as
\beq \label{eq:sph_sym_expansion}
\Delta\beta = \Delta\beta_0\e^{\iu\phi}, \ \Delta\gamma = \Delta\gamma_0\e^{2\iu\phi} , \ F = F_0\e^{\iu\phi} , \  G = G_0\e^{3\iu\phi} ,
\eeq 
where $ \Delta\beta_0, \Delta\gamma_0, F_0, G_0 $ are real numbers that depend on $h$ but not on $\phi$. Note that $ \Delta\kappa $ is independent on $ \phi  $. 

Axisymmetry is mildly broken in realistic elliptic profiles; we discuss this point in App.~\ref{app:ellitp_nfw}. However, as long as the ellipticity is small, the $ \phi $ dependence of $ \Delta\beta_0, \Delta\gamma_0, F_0, G_0 $ is weak, and does not affect our main point.   

Of the expansion coefficients in Eq.~(\ref{eq:aexp}), $\Delta\beta$ is degenerate with $\beta$, which is a free parameter in the modeling, and has no effect on time delays. $\Delta\kappa$ is degenerate with $\kappa_{\rm ext}$, a free parameter.\footnote{A cosmological prior can sometimes be imposed on $\kappa_{\rm ext}$, by comparing the system's field to N-body simulations.} Varied along the MSD, the combination $\Delta\kappa+\kappa_{\rm ext}$ is invisible to imaging, but affects time delays. $\Delta\gamma$ is degenerate with $\gamma_{\rm ext}$, a free parameter. Thus, as far as imaging data is concerned, the leading order effect that constrains the group comes from the MSD-invariant combinations $F/(1-\Delta\kappa - \kappa_{\rm ext})$ and $G/(1-\Delta\kappa - \kappa_{\rm ext})$, related to what is known in the literature as reduced flexion~\cite{Goldberg:2004hh,Bacon:2005qr}. 

An attempt to constrain the group model via imaging data, suffers from the following three main difficulties. The first obvious point is that the flexion terms are small; in the $1/h$ expansion, the flexion terms are parametrically suppressed as $F,G\propto \Delta\kappa/h$.  The second point is that direct modeling of the group halo can still leave room for a residual MSD: a model to capture correctly the imaging distortion produced by the flexion, while mismodeling the convergence. The third point, which may be the most important in practice, is that the posterior likelihood of the group halo model exhibits a discrete approximate degeneracy, related to the $2\pi/3$ phase degeneracy of the $G$ term (see Eq.~(\ref{eq:sph_sym_expansion})).

It is useful to study a simple example. 
Consider a group halo described by an isotropic power-law (PL) density profile with 3D slope $\gamma_{\rm g}$ and Einstein angle $\theta_{\rm Eg}$. In this case, the coefficients of Eqs.~\eqref{eq:aexp} and~\eqref{eq:sph_sym_expansion} are given by
\be
\label{eq:dkPL}\Delta\kappa&=&\frac{3-\gamma_{\rm g}}{2}\left(\frac{\theta_{\rm Eg}}{h}\right)^{\gamma_{\rm g}-1},\;\;\;
\label{eq:dgPL}\Delta\gamma_0=\frac{1-\gamma_{\rm g}}{3-\gamma_{\rm g}}\Delta\kappa,\\
\label{eq:dfgPL}F_0&=&\left(\gamma_{\rm g}-1\right)\frac{\Delta\kappa}{h},\;\;\;
G_0=-\frac{\gamma_{\rm g}^2-1}{3-\gamma_{\rm g}}\frac{\Delta\kappa}{h},\\
\label{eq:dbPL}\Delta\beta&=&-\frac{2\,h}{3-\gamma_{\rm g}}\Delta\kappa.\ee

For $\gamma_{\rm g}\approx2$ (singular isothermal sphere (SIS)), the line of sight velocity dispersion (LOSVD) in the center of the group is related to the group's Einstein radius via\footnote{See App.~\ref{app:kin} for a review of the derivation. For $\gamma_{\rm g}$ close, but not equal to 2, the RHS of Eq.~(\ref{eq:tEgs2}) is rescaled by an $\mathcal{O}(1)$ factor, equal to $\sim0.5$ ($\sim1.5$) for $\gamma_{\rm g}=1.5$ ($\gamma_{\rm g}=2.2$), and a weak dependence arises on the scale radius of the profile. 
}
\be\label{eq:tEgs2}\theta_{\rm Eg}&\approx&\frac{4\pi d_{\rm ls}}{d_{\rm s}}\sigma^2_{\rm los}\,\approx\,4.5''\,\frac{d_{\rm ls}}{d_{\rm s}}\left(\frac{\sigma_{\rm los}}{400~\rm km/s}\right)^2.\ee
We can therefore estimate the convergence,
\be\Delta\kappa_{\rm SIS}&\approx&0.5\frac{d_{\rm ls}}{d_{\rm s}}\left(\frac{\sigma_{\rm los}}{400~\rm km/s}\right)^2\left(\frac{20''}{h}\right),\ee
and to the flexion terms,
\beq F_{0,\rm SIS}\,\approx\,\frac{0.025}{1''}\left(\frac{20''}{h}\right)\Delta\kappa_{\rm SIS},\;\;\;
G_{0,\rm SIS}\,\approx\,-3\,F_{0,\rm SIS}.\eeq
For example, $\Delta\kappa\approx0.1$ produced by a SIS group with $d_{\rm ls}/d_{\rm s}=0.5$, $\sigma_{\rm los}=400$~km/s, and $h=50''$, would cause (if not modeled) a $\sim 10\%$ bias in the inference of $H_0$, while the imaging distortion produced by the group's flexion field at $\theta\sim1''$ would only amount to $\delta\theta\approx0.0075''$, typically dominated by the $G$ term. Next order terms in the expansion (beyond the flexion) are further suppressed by $\sim1''/h\sim\mathcal{O}(1/10)$, and can be neglected.

This discussion highlights the obvious hierarchy between flexion and convergence, but as was mentioned earlier, there is also MSD. In Eqs.~(\ref{eq:dkPL}-\ref{eq:dfgPL}), the leading effect of the MSD (at small $\Delta\kappa\ll1$) is seen by noticing that even if one can fix $F$ and $G$ exactly from imaging data, this still allows $\Delta\kappa$ to vary freely, as long as $h$ is varied simultaneously with $\Delta\kappa\propto h$. Thus, unless we have good external prior data on the group's center of mass (parameterised by $h$), determining the flexion from imaging data alone does not fix the convergence. (More precisely, the degeneracy is somewhat regulated by the fact that the true MSD invariant quantities are $F/(1-\Delta\kappa - \kappa_{\rm ext})$ and $G/(1-\Delta\kappa - \kappa_{\rm ext})$, rather than $F$ and $G$ themselves. The exact MSD is then captured via the freedom to adjust $\Delta\kappa$ and $h$ while keeping $(\Delta\kappa/h)/\left(1-\kappa_{\rm ext}-\Delta\kappa\right)$ constant.)

Finally, another important point is the phase degeneracy of the $G$ term. This degeneracy means that models of the halo in which the direction to the group's center is changed by $\phi\to\phi\pm2\pi/3$ yield identical $G$ terms. Although the rotated models do not reproduce the truth value of the $F$ term, the $G$ degeneracy can produce isolated local minima in the posterior likelihood, that can trap unweary MCMC chains. Interestingly, the offset $F$ term in the ``wrong" minima causes the model to  pull towards a biased estimate of the group's convergence, thereby biasing $H_0$. This issue is seen to be an important point in the next section.

\section{Illustration with mock data}\label{s:mock}
The host group is often modeled explicitly if the system is known to reside in a group (see e.g.~\cite{Millon:2019slk,Momcheva:2005ex,Wilson:2016hcs,Wilson:2017apg}). 
We now study such modeling using mock data. Our implementation is based on the package \texttt{lenstronomy}~\cite{Birrer:2018,Birrer:2015rpa,Birrer:2021wjl}.

\subsection{Mock setup description}
\begin{figure}
	\centering
	\includegraphics[scale=0.35]{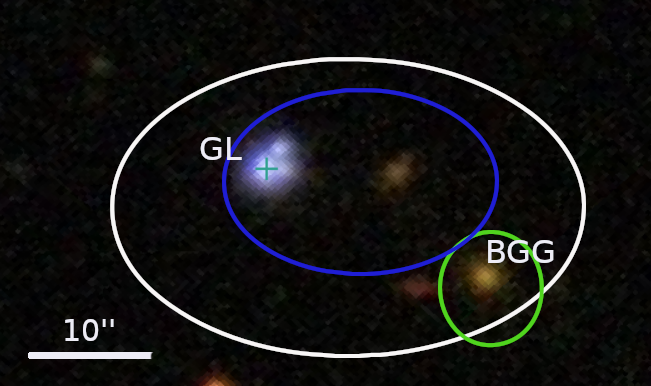}
	\caption{The field of PG1115+080 (image from SIMBAD~\cite{Wenger:2000sw}). The primary lens, GL, is marked by a cross, surrounded by the quasar images. White ellipse shows the 68\% CL prior derived in~\cite{Wilson:2016hcs} and used in~\cite{Chen:2019ejq} to constrain the group's center of mass. Green and blue ellipses give a rough illustration of the X-ray emission associated to the group, found in~\cite{Grant:2003yw} and~\cite{Fassnacht:2007xk}, respectively, using different subtraction schemes for the quasar emission. 
	}
	\label{fig:HSTXray}
\end{figure}

We chose PG1115+080 as a reference object to guide our study. Ref.~\cite{Millon:2019slk} inferred $H_0=81^{+8}_{-7}$~km/s/Mpc from this system, quite high compared to the CMB value~\cite{Akrami:2018vks}. At the same time, the system is known to reside in a group~\cite{Wilson:2017apg} with  
$\sigma_{\rm los}=390^{+50}_{-60}~{\rm km/s}$ (an earlier study found $\sigma_{\rm los}=440^{+90}_{-80}~{\rm km/s}$~\cite{Momcheva:2005ex}), estimated to induce $\kappa_{\rm g}\sim0.2$. 
The group's projected center of mass is not far from the primary lens ($h$ of the order of $10''$). This suggests that flexion terms are probably not negligible for this system, distorting the image on scales $\Delta\theta\sim\kappa_{\rm g}\theta_{\rm E}^2/h\sim0.01''$. 
An image of the field is shown in Fig.~\ref{fig:HSTXray}.

\begin{figure}
    \centering
    \includegraphics[scale=0.5]{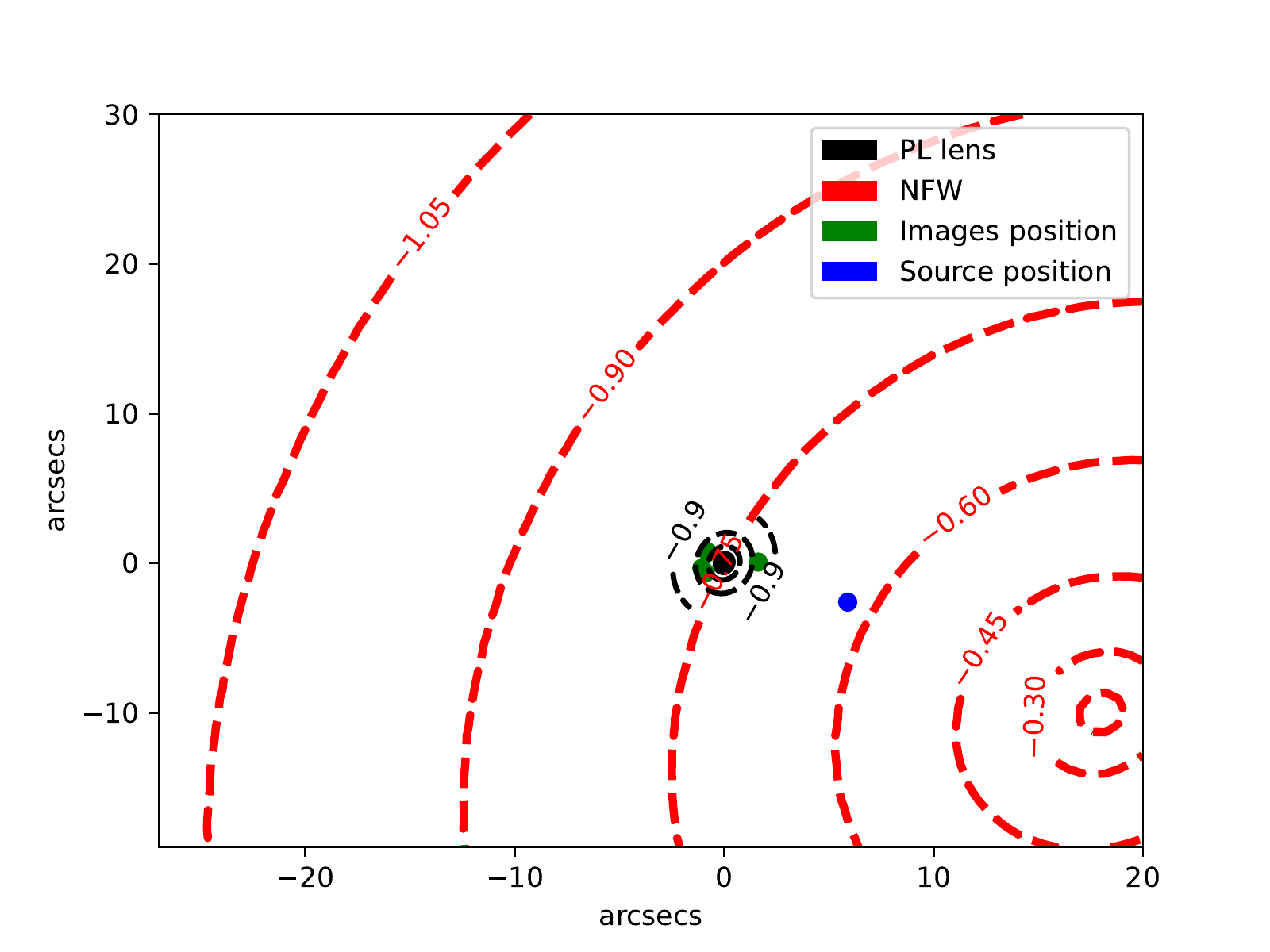}
    \caption{Illustration of the mock setup. Contours show $ \log_{10} $ of the convergence for the primary lens and the NFW halo in black and red, respectively \href{https://github.com/lucateo/Group_Halo_Degeneracy/blob/main/mock_setup.ipynb}{\faGithub}.}
    \label{fig:mock}
\end{figure}

We consider the following mock setup.
For the primary lens, we consider an elliptic power-law density profile with 3D slope $\gamma=2.17$, Einstein angle $\theta_{\rm E}=1.08''$, and ellipticity parameters $e_1=-0.2,\,e_2=0.05$, corresponding to $q=(1-e)/(1+e)\approx0.66$, where $e=\sqrt{e^2_1+e^2_2}$. 
For the group, we consider an elliptic NFW profile~\cite{Jing:2002np,Vega-ferrero:2016xoa} with $ e_1 = -0.07 $, $ e_2 = 0.03 $, compatible with the findings of~\cite{Schrabback:2020bsp}.
Fig.~\ref{fig:mock} illustrates the setup, including the truth position of the group halo center of mass. This setup could mimic PG1115+080 (Fig.~\ref{fig:HSTXray}) if the BGG -- or the X-ray blob found by~\cite{Grant:2003yw} -- happens to indicate the group's center of mass position. 

%

Following~\cite{Chen:2019ejq}, we choose our inference pipeline to include only a simple spherical group model, ignoring group halo ellipticity in the modeling.

\subsection{External priors: tracer galaxy kinematics and theoretical input from N-body simulations}\label{ss:extprior}
As in~\cite{Chen:2019ejq}, we include external priors on the group halo, dictated by cosmological N-body simulations and by kinematics data. We defer most of the details to App.~\ref{app:kin}, notably Secs.~\ref{subsec:NFW_kinematics} and~\ref{ss:kinprior}. 
However, we would like to point out a potential inaccuracy in the kinematics analyses of some previous works.  

Galaxy groups are often assumed to follow the NFW density profile~\cite{Navarro:1996gj},
\be\label{eq:rhpNFW}
	\rho (r) = \frac{\rho_0 R^3_{\rm s}}{r(r+R_{\rm s})^2}  ,
\ee 
In terms of the parameters $\rho_0$ and $R_{\rm s}$, the LOSVD of tracer galaxies (with number density assumed to follow the same profile as the dark matter density) can be expressed as
\be\label{eq:losvdNFW0}\sigma_{\rm los}^2(\theta)&=&G\rho_0R_{\rm s}^2\,f\left(\frac{\theta}{\theta_{\rm s}}\right),
\ee
where $f(a)$ is a dimensionless function derived in Sec.~\ref{subsec:NFW_kinematics}.

While Eq.~(\ref{eq:losvdNFW0}) (averaged as needed within some aperture cut of the observations) gives the correct translation between LOSVD data and the NFW model parameters, Ref.~\cite{Chen:2019ejq} (after~\cite{Koranyi:1999jz,Wong:2010xk}) considered a different expression as a proxy for the LOSVD data:
\be \label{eq:sigma0}
\bar\sigma^2&=&\frac{GM_{\rm vir}}{3R_{\rm vir}},\ee
where $R_{\rm vir}$ and $M_{\rm vir}=M(R_{\rm vir})$ are the virial radius and virial mass, and $c_{\rm vir}=R_{\rm vir}/R_s$ is the  NFW concentration parameter.
In App.~\ref{app:kin} we show that identifying the quantity $\bar\sigma^2$ with the observable $\sigma^2_{\rm los}$ introduces an error of up to $\sim20\%$ (the precise error depends on the analysis aperture). 

One reason for introducing the auxiliary quantities $R_{\rm vir}$ and $M_{\rm vir}$, is that  N-body simulations provide theoretically-motivated priors that are often presented in terms of these quantities~\cite{Eke:2000av,Maccio:2008pcd}. These cosmological priors are, however, quantitatively and conceptually decoupled from the kinematics data interpretation. Instead, as we review in App.~\ref{app:kin}, the cosmological priors dictate a certain redshift-dependent relation between the NFW parameters $\rho_0$ and $R_{\rm s}$. There is no obstacle to implement this theoretical prior while still maintaining the correct kinematical expression, Eq.~(\ref{eq:losvdNFW0}).

In our main implementation of the MCMC, we set the standard deviation for $\sigma_{\rm los}$ to $120$~km/s. This doubles the nominal uncertainty quoted by~\cite{Wilson:2017apg} for the group of PG1115+080, but we believe that such cautionary procedure is reasonable. 
To be clear, we are not suggesting to doubt the observational LOSVD from~\cite{Wilson:2017apg}. Rather, the uncertainties we worry about concern the theoretical interpretation within simplified halo models. 
The systematic error due to using $\bar\sigma$ instead of $\sigma_{\rm los}$, as done in~\cite{Chen:2019ejq}, is of the order of $\delta\sigma_{\rm los}^2/\sigma_{\rm los}^2\sim20\%$, so $\delta\sigma_{\rm los}/\sigma_{\rm los}\sim10\%$, or $\delta\sigma_{\rm los}\sim40$~km/s, quite comparable to the ``bare" observational uncertainty. There are  additional plausible errors: the group is likely to be aspherical~\cite{Schrabback:2020bsp}, the velocity distribution need not be isotropic, and the group may not be fully virialised. Each of these could cause systematic shifts of tens of percent in the kinematics interpretation. 

For completeness, in App.~\ref{app:1} we check how a LOSVD uncertainty of $60$~km/s changes the results. We find the difference to be quantitatively insignificant for our main results.

\subsection{Results}\label{ss:results}
First, to obtain a global view of the posterior ``landscape", we run the \texttt{zeus} MCMC algorithm~\cite{Karamanis:2020zss,Karamanis:2021tsx}, which is designed to cope with multiple likelihood minima. 
\begin{figure*}
    \centering
  \includegraphics[scale=0.4]{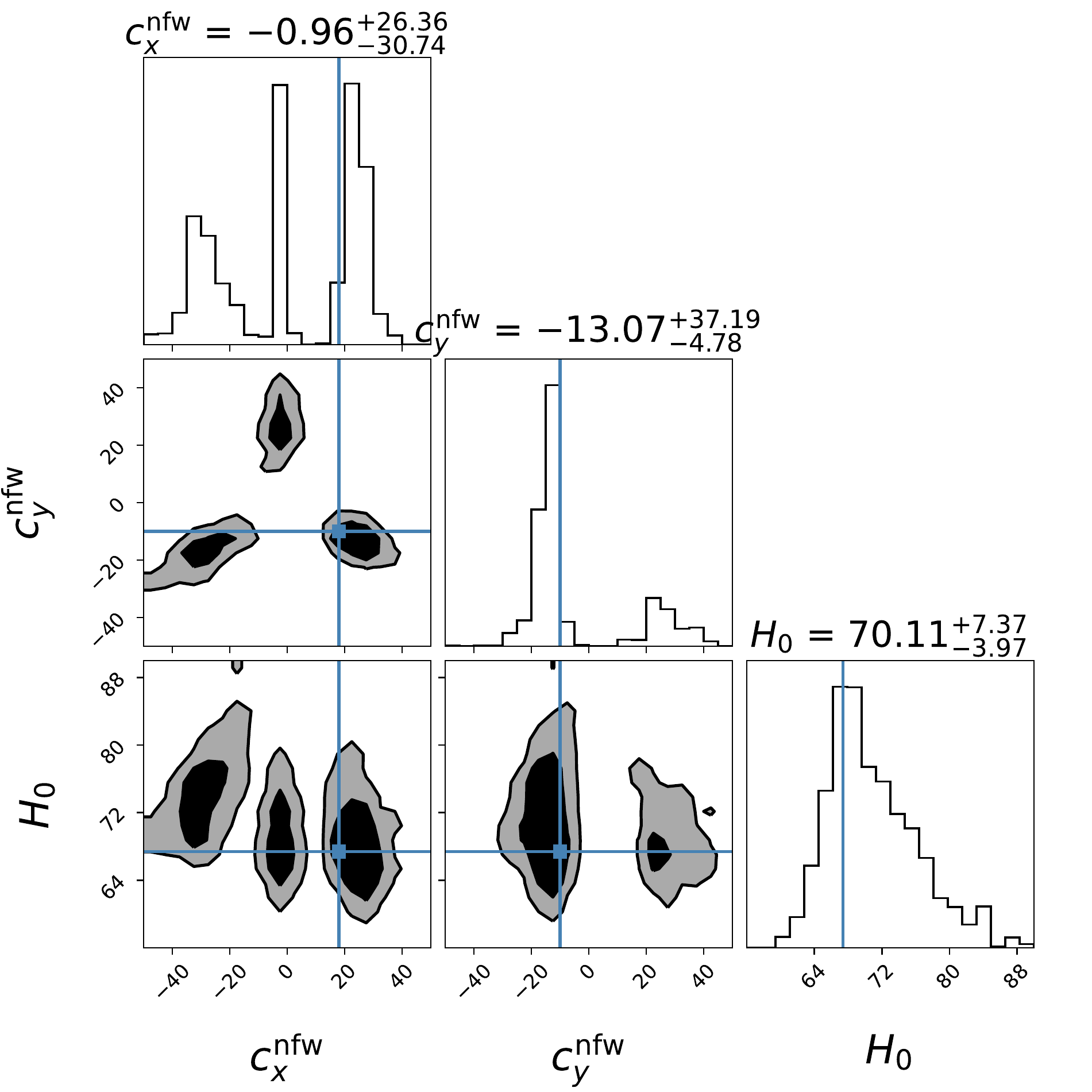}
    \includegraphics[scale=0.4]{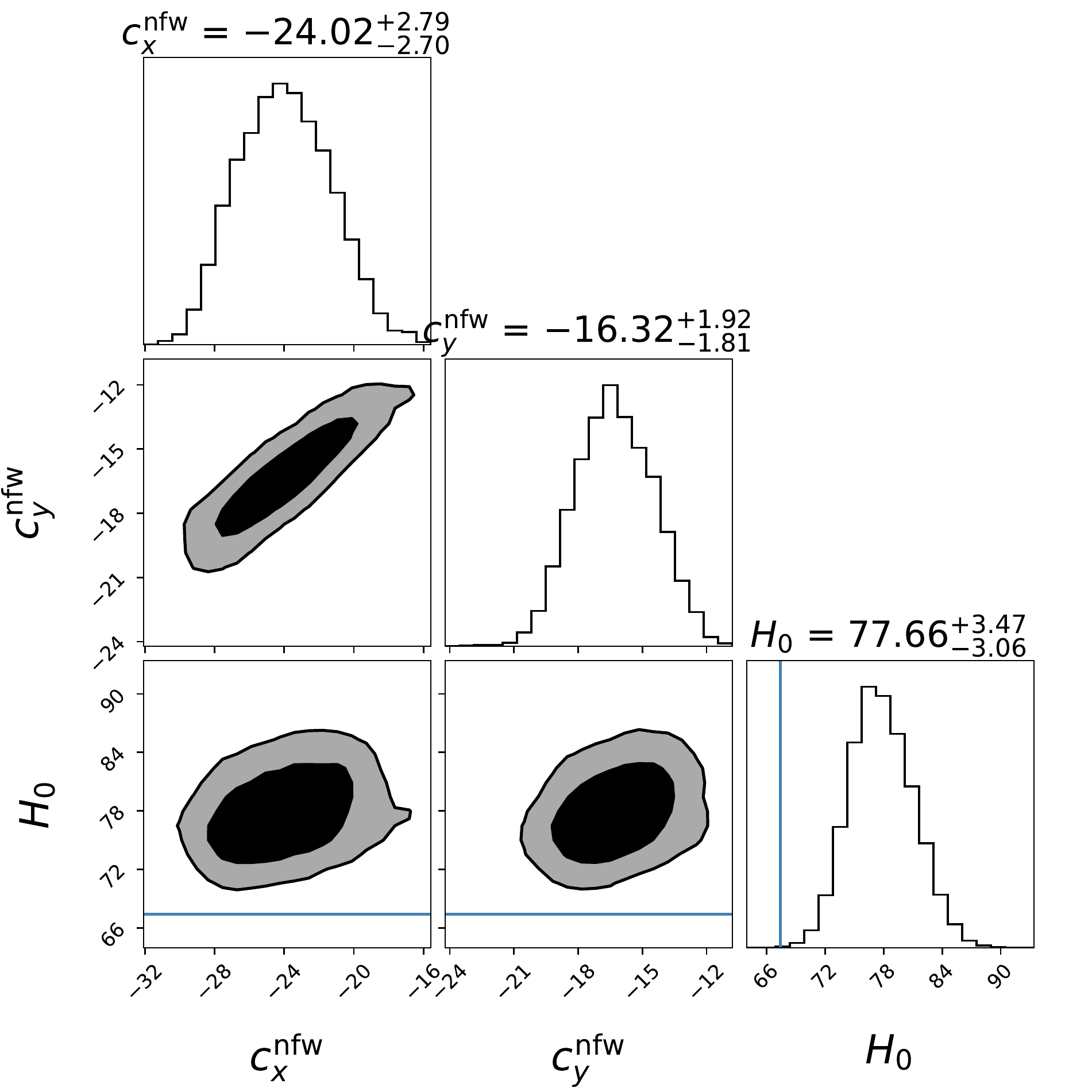}
    \includegraphics[scale=0.4]{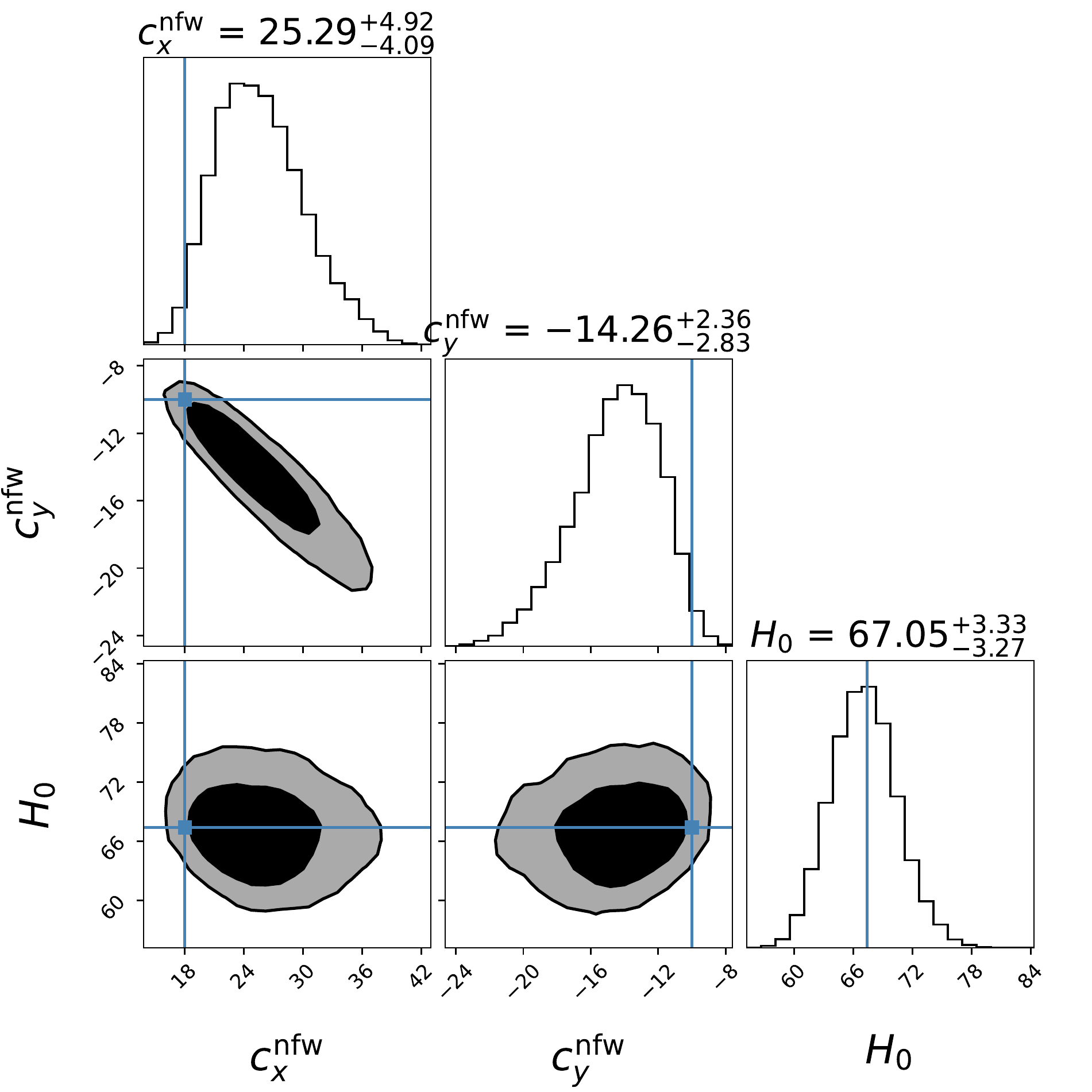}
    \includegraphics[scale=0.4]{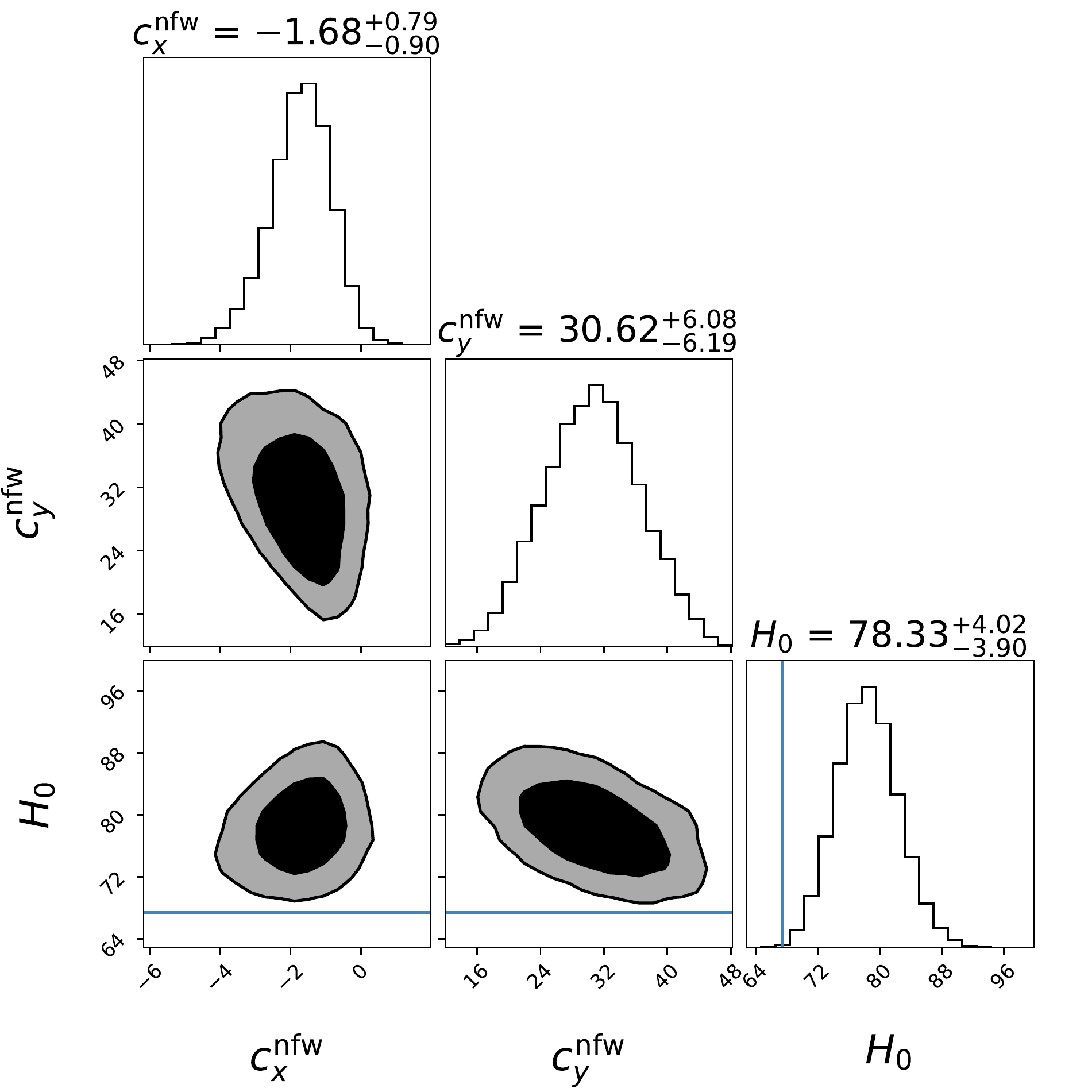}
    \caption{Mock analysis. Truth set-up: elliptic group located at $c_x^{\rm nfw}=18''$, $c_y^{\rm nfw}=-10''$. {\bf Top left panel}: \texttt{zeus} run, exposing the global likelihood landscape. {\bf Other panels}: \texttt{emcee} runs, falling into different local minima. The \texttt{emcee} runs are initiated with different priors for the position of the  group center, keeping the same Gaussian standard deviation of $16'' $. See also Fig.~\ref{fig:mock_inference} \href{https://github.com/lucateo/Group_Halo_Degeneracy/tree/main}{\faGithub}.
    }
    \label{fig:mins}
\end{figure*}
We show the main result in the {\bf top-left panel} of Fig.~\ref{fig:mins}. 
Notice the 3 disjoint minima in the posterior likelihood in the $c_x^{\rm nfw}-c_y^{\rm nfw}$ panel, where $c_{x,y}^{\rm nfw}$ are the angular coordinates of the halo center of mass. The origin of this threefold degeneracy is the $\phi\to\phi\pm2\pi/3$ degeneracy of the flexion $G$ term. This run has no prior on $c_x^{\rm nfw}$ and $c_y^{\rm nfw}$. More details and comprehensive MCMC corner plots are provided in App.~\ref{app:1}. 

\begin{figure}
	\centering
	\includegraphics[width=0.475\textwidth]{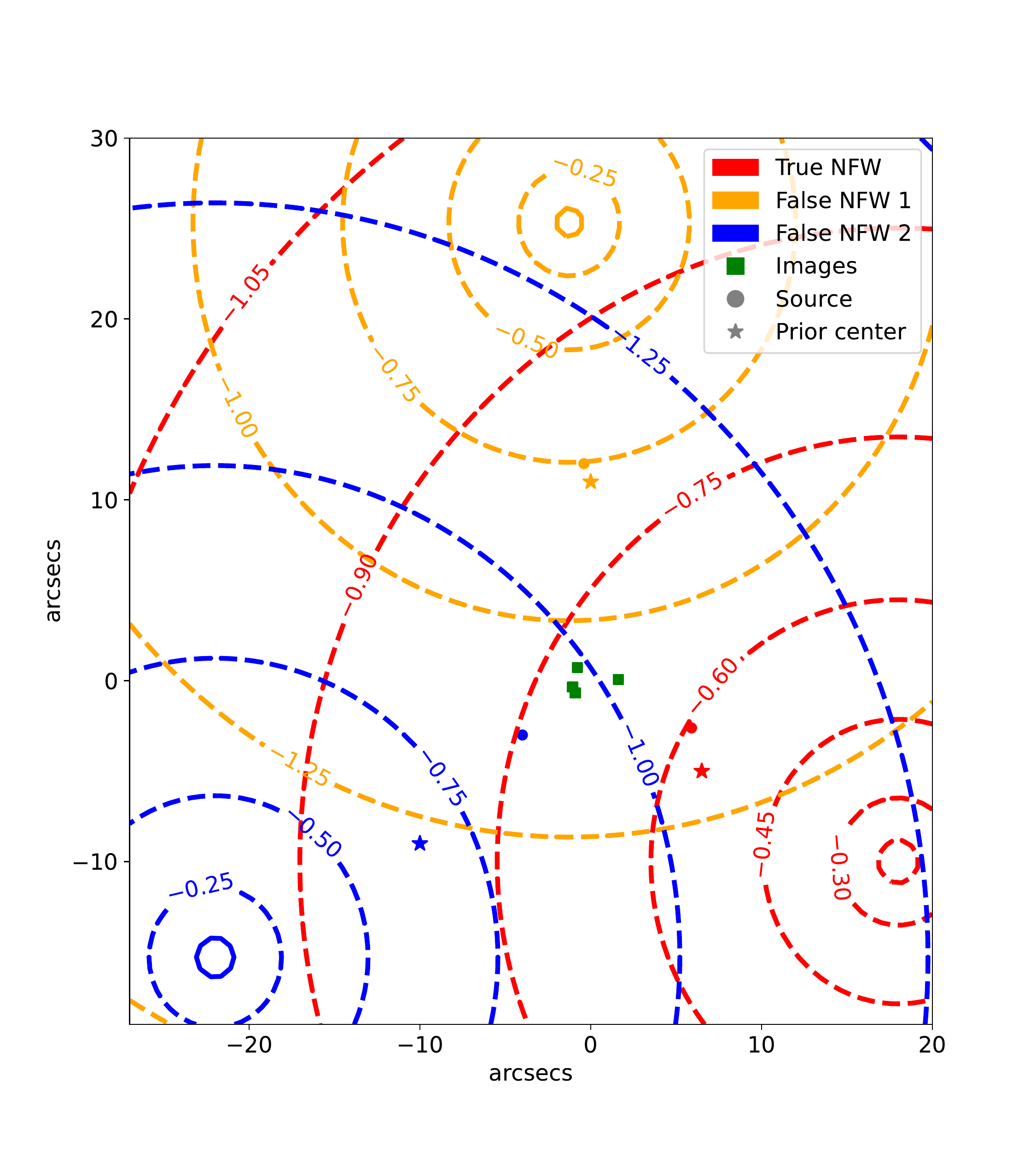}
	\caption{Illustration of the three posterior likelihood minima probed by the MCMC inference. Contours show $ \log_{10} $ convergence isocontours, for the 3 group halo solutions (red, orange, and blue, respectively). Dots and stars highlight the inferred source position and the center of the group prior, respectively. Green squares denote the position of the images. The parameter values used in the plot are read from the \texttt{emcee} run best fit, see Fig.~\ref{fig:mins} \href{https://github.com/lucateo/Group_Halo_Degeneracy/blob/main/mock_setup.ipynb}{\faGithub}.}
	\label{fig:mock_inference}
\end{figure}

The local minima could trap an MCMC, if the scanning algorithm is not suited to probe multimodal posteriors.
To see this, we repeat the analysis, this time using the \texttt{emcee} algorithm~\cite{Foreman_Mackey:2013}. The results are shown in the {\bf top-right} and {\bf bottom panels} of Fig.~\ref{fig:mins}. Indeed, \texttt{emcee} tends to discover only one of the local minima, missing the others.
In the {\bf top-right} and {\bf bottom-right panels}, \texttt{emcee} converges on a biased group halo position. The biased local minima yield an $H_0$ posterior that is on the high side, pulling $H_0$ above its truth value by $\sim3\sigma$ and $\sim1.8\sigma$, respectively. The convergence pattern associated to the three minima is shown in Fig.~\ref{fig:mock_inference}.

Performing more MCMC runs, we find that \texttt{emcee} consistently converges into just one of the minima, missing the others.  
The choice of the minimum found by \texttt{emcee} mostly depends on the initial position of the MCMC walkers in the $ c_x^{\rm nfw}-c_y^{\rm nfw} $ parameter space. In our runs, the initial walker allocation is guided by a Gaussian prior with a similar 68\%CL radius as that used in~\cite{Chen:2019ejq} (see Fig.~\ref{fig:HSTXray}), and with centers shown by star symbols in Fig.~\ref{fig:mock_inference}. 

After performing a series of numerical trials we emphasize that, at least for the given, rather broad (but realistic) width of the prior, the starting point of the MCMC walkers appears to be a more significant factor in determining which of the three minima traps the chain, than the prior center itself. These considerations coincide if the MCMC walkers are initiated at the prior center. 

Given the above discussion, we can make a rough estimate of the probability of the MCMC to fall into a displaced minimum, by matching it with the probability of the group center prior to be nearer a false minimum than the truth one. In Sec.~\ref{ss:probfalse} we estimate this probability from mock realizations of samples of tracer galaxies. The result depends on the analyses aperture, the number of galaxies in a sample, and the underlying group profile. For 13 members in the NFW profile, with an aperture of $\sim10R_s$ (i.e., around 3 virial radii), the false minimum probability we find is $\sim10\%$.

We comment that X-ray data may help to pinpoint the centroid of a massive group, resolving the threefold degeneracy. For example, Ref.~\cite{Grant:2003yw} found an X-ray blob that appears to be centered around the BGG of the host group of PG1115+080 (green contour in Fig.~\ref{fig:HSTXray}). 
If the X-ray emission can be associated with the group's centroid, it may provide a narrower prior than that derived from tracer galaxies. 
The X-ray analysis may be complicated by blending with the lensed quasar: using a different method to mask the quasar, Ref.~\cite{Fassnacht:2007xk} found a shifted, more extended, and brighter group emission. Even so, it seems plausible that X-ray data could help narrowing the group's centroid prior\footnote{In principle, X-ray luminosity measurements can also constrain the group's velocity dispersion and mass. In practice, however, the limited accuracy of X-ray luminosity measurements~\cite{Fassnacht:2007xk}, combined with the observed scatter in luminosity--mass or luminosity--dispersion correlations~\cite{Lovisari:2021lmb}, suggests that the corresponding constraints may not lead to significant improvement over kinematics alone.}.

It is natural to ask whether the discrete modeling bias tends to come along with an anomalously large external shear estimate. While this could indeed happen, and might serve as a useful ``alarm bell" if the $H_0$ bias is larger than 10\% or so, our numerical examples also show false minima solutions with acceptable values of external shear. For example, Fig.~\ref{fig:nocprior_false_minimum_full} shows a false minimum solution with $\sim$15\% bias in $H_0$ but with external shear values that are compatible with those found in~\cite{Chen:2019ejq} for PG1115+080 (see Figs.~7,8 there).

Finally, we emphasize again that PG1115+080 was selected to guide our mock specifically because of its massive, near-by group association. This set-up, while we believe it deserves study on its own right, may be an un-representative outlier among lensed quasar systems, and it is therefore natural to ask just how uncommon it is. 
A detailed analysis of the fraction of systems that may exhibit flexion degeneracy is beyond the scope of this work, but in App.~\ref{app:d} we attempt a crude estimate using the data from~\cite{Wilson:2017apg}. Our results suggest that detectable flexion distortion from LOS groups (including, not limited to, the host of the primary lens) may affect $\sim10\%-30\%$ of lensed quasar systems.

\section{Origin of $H_0$ bias in a displaced group} \label{s:Bias}
As we explained, the three-fold approximate degeneracy manifest in Fig.~\ref{fig:mins} is due to the behavior under rotations of the $ G $ flexion. The $ F $ flexion, however, behaves like a vector. Hence, when the inference falls into a wrong minimum, it attempts to minimize  the difference between the truth $F$ deflection and the wrong-minimum inference value of $F$. This can be achieved by keeping $ G_{\rm eff} $ the same, but reducing 
$ F_{\rm eff} $ as much as possible, where 
\begin{equation} \label{eq:reducedFlexion}
F_{\rm eff} = \frac{F_0}{1-\Delta \kappa} \ , \ G_{\rm eff} = \frac{G_0}{1-\Delta \kappa}  
\end{equation}
are the reduced flexion terms (for clarity, here we omit the cosmological term $\kappa_{\rm ext}$).
To see this point, denote the reduced flexion of the inference model by $ F_{\rm eff} $, and denote the truth flexion by $F^{(\rm truth)}_{\rm eff}$.\footnote{The truth and model values of the $G$ term are assumed to approximately coincide, $G^{(\rm truth)}_{\rm eff}\approx G_{\rm eff}$.} We  expect the MCMC to minimize
\be \label{eq:minimizeF}
&\abs{F_{\rm eff}^{(\rm truth)}  - F_{\rm eff} \e^{\frac{2\iu n\pi}{3}}} =\no\\
&\left|F^{(\rm truth)}_{\rm eff}\right| \sqrt{1-2\frac{F_{\rm eff}}{F^{(\rm truth)}_{\rm eff}}\cos\left(\frac{2n\pi}{3}\right)+ \left(\frac{F_{\rm eff}}{F^{(\rm truth)}_{\rm eff}}\right)^2 } ,\;\;
\ee 
with $ n=\pm1$ selecting the position of the false minimum. Since $ \cos(\pm2\pi/3) = -1/2 $, the above expression is minimized for $ F_{\rm eff} = 0 $. 

Consider the PL model of Sec.~\ref{s:group}. In this model, we have $F_{\rm eff}/G_{\rm eff}=(\gamma_{\rm g}-3)/(\gamma_{\rm g}+1)$. 
Therefore, for $1<\gamma_{\rm g}<3$ (the range of interest), decreasing $|F_{\rm eff}|$ at fixed $G_{\rm eff}$ entails increasing $\gamma_{\rm g}$, while adjusting the other parameters of the model so as to keep $G_{\rm eff}$ constant. Those other parameters were introduced in Eqs.~(\ref{eq:dkPL}-\ref{eq:dfgPL}) as $h$ and $\theta_{\rm Eg}$, but we can equally well replace $\theta_{\rm Eg}$ by $\Delta\kappa$. Now, we have 
$G_{\rm eff}=-\frac{\gamma^2_{\rm g}-1}{3-\gamma_{\rm g}}\frac{\Delta\kappa}{h\left(1-\Delta\kappa\right)}$. The $\gamma_{\rm g}$-dependent factor, $\frac{\gamma^2_{\rm g}-1}{3-\gamma_{\rm g}}$, increases with increasing $\gamma_{\rm g}$; to compensate for this and keep $G_{\rm eff}$ constant, the factor $\frac{\Delta\kappa}{h\left(1-\Delta\kappa\right)}$ needs to decrease. For small $\Delta\kappa\ll1$, this means that near any one of the displaced likelihood minima, the MCMC will attempt to decrease $\Delta\kappa/h$ in comparison to its truth value. Part of this adjustment entails decreasing the model value of $\Delta\kappa$, which therefore biases $H_0$ high, as $\Delta H_0/H_0\approx-\left(\Delta\kappa^{(\rm truth)}-\Delta\kappa^{(\rm model)}\right)$. 

In App.~\ref{app:flexion} we show that a similar analysis holds also for the NFW model, used in the MCMC implementation: also in that case, falling into a displaced group minimum causes the fit to underestimate of $\Delta\kappa$, leading to an overestimate of $H_0$. This analysis clarifies the trend seen in Fig.~\ref{fig:mins}. 

\section{Summary}\label{s:sum}
Lens galaxies in quasar lensing time delay measurements are often members of galaxy groups, that must be modeled for an accurate determination of $H_0$. The group modeling exhibits approximate versions of the MSD (Sec.~\ref{s:group}). Essentially, it is a displaced-center version of the core-MSD considered in~\cite{Schneider_2013,Blum:2020mgu}. 

At leading order in the tidal approximation, the group halo enters imaging through the flexion. We showed an approximate threefold discrete modeling degeneracy, associated with rotating the assumed position of the group centroid by an angle of $2\pi/3$ around the primary lens (Secs.~\ref{s:group} and ~\ref{s:mock}). This produces a posterior likelihood with three disjoint minima. MCMC algorithms that fail to expose this structure may fall into a displaced minimum. The inferred value of $H_0$ found in a displaced minimum is systematically biased high (Sec.~\ref{s:Bias}). Using numerical mock data experiments motivated by a realistic system, we demonstrated that the $H_0$ bias can reach $\sim10\%$.

The choice of the minimum detected by the MCMC can strongly depend on the starting position of the walkers in the space of group centroid coordinates. If the starting point is chosen as the centroid prior center, then the probability for the MCMC to land in a displaced minimum may be rather small. For a sample of 13 tracer galaxies (relevant for PG1115+080) with an aperture of about 3 virial radii in an NFW halo, this probability is $\sim10\%$. 


Our analysis suggests the following recommendations.
\begin{enumerate}
\item Bayesian cosmography analyses should explore the full posterior likelihood landscape. Awareness of this possible three disjoint minima structure, if not already there, is needed. 
%
%
%
\item X-ray data may help to pinpoint the centroid of a massive group, resolving the  degeneracy. 
\item As an aside, we note that some cosmography analyses (e.g.~\cite{Chen:2019ejq}) used an incorrect kinematics prior to constrain the group model. The error in the interpretation of tracer galaxy velocity dispersion depends on the analysis aperture, and can reach $\sim20\%$. A correct version of the kinematics prior is reviewed in App.~\ref{app:kin}.  
\end{enumerate}

The number of strong lensing time delay systems is expected to increase by more than an order of magnitude in the near future~\cite{Oguri:2010ns,Liao:2014cka,Dobke:2009bz}, an important step towards possibly reaching a few percent lensing determination of $H_0$~\cite{Birrer:2020jyr}. This program could be further assisted by many  resources~\cite{DES:2018whv,SKA:2018ckk,Euclid:2019clj,LSSTScience:2009jmu,Gardner:2006ky}. Our study highlights some pitfalls (and suggests solutions) that need to be taken into account if the precision goal for $H_0$ should also be accurate.

\acknowledgments
We thank Simon Birrer, Marko Simonović, and Raphael Flauger for useful discussions. 
This work made use of the following public software packages: \texttt{lenstronomy} \cite{Birrer:2018,Birrer:2015,Birrer:2021wjl}, \texttt{emcee} \cite{Foreman_Mackey:2013}, \texttt{zeus}~\cite{Karamanis:2020zss,Karamanis:2021tsx}, \texttt{corner} \cite{Foreman-Mackey2016}, \texttt{astropy} \cite{Robitaille:2013mpa,Price-Whelan:2018hus}.
This research has made use of the SIMBAD database,
operated at CDS, Strasbourg, France~\cite{Wenger:2000sw}.
This work was supported by the Israel Science Foundation grant 1784/20, and by MINERVA grant 714123. LT wishes to acknowledge association with the International Helmholtz-Weizmann Research School for Multimessenger Astronomy.

%

\begin{appendix}
\section{Deflection angle expansion, NFW profile} \label{app:flexion}	
We start with some general preliminaries; for the discussion of the NFW profile, the reader can skip to Eq.~(\ref{eq:rhpNFW}).

For our purpose, which is to analyze the lensing equation in the vicinity of a particular galaxy member of the group, it is convenient to use a coordinate system that is centered on the primary lens galaxy, and displaced from the group center of mass by a separation angle $\vec h$. 
In these coordinates the group lensing potential reads
\be
	\Psi_{\rm halo}(\theta_1, \theta_2) := \Psi_{\rm halo}(|\vec\theta - \vec{h}|) . 
\ee
We can expand with respect to the small parameters $ \theta_1/h, \ \theta_2/h $,
\beq \label{eq:expansion}
 \Psi_{\rm halo}(\theta_1, \theta_2) \simeq\Psi_{\rm halo}(0, 0) + \grad_{\vec\theta} \Psi_{\rm halo}(0, 0) \cdot 	\vec\theta + \ldots \ .
\eeq
Recall that the lensing potential and the deflection angle are related as $ \grad\Psi = \vec{\alpha} $; in particular, we can write the lens equation as
\begin{equation}
	\beta_i = \theta_i - \partial_i \Psi_{\rm lens}(\vec\theta) - \partial_i \Psi_{\rm halo}(0,0) - \partial_i\partial_j \Psi_{\rm halo}(0,0) \theta_j + \ldots \ .
\end{equation} 
As discussed on the main text, in complex notation $ \vec{h} \to h\e^{\iu\phi} $. The lensing potential $ \Psi $ in this formalism is real. One can obtain the complex deflection angle by means of the derivative operator
\begin{equation}
	\grad_{\rm c} := \partial_1 + \iu\partial_2 \implies \alpha = \grad_{\rm c} \Psi \ .
\end{equation} 
In the expansion of the lensing potential, it is easy to see that
\begin{equation}
\Delta\beta=\grad_{\rm c}\Psi_{\rm halo} , \	\Delta\kappa = \frac{1}{2} \grad_{\rm c}\grad^*_{\rm c} \Psi_{\rm halo} , \ \Delta\gamma =  \frac{1}{2}\grad_{\rm c} \grad_{\rm c} \Psi_{\rm halo} ,
\end{equation}
where the derivatives are computed at the origin.
Analogously, we can express third order derivatives of the lensing potential as derivatives of the shear,
\begin{equation}
	G = \grad_{\rm c}\Delta\gamma , \ F = \grad^*_{\rm c}\Delta\gamma  ,
\end{equation}
where $ G $, $ F $ are the flexion terms. 

In the axisymmetric case, a group halo center at $ \phi=0 $ and fixed $ h $ would yield the same expansion coefficients of a halo located at a generic $ \phi $, granted that we rotate our coordinate system accordingly. 
We can pass to cylindrical coordinates\footnote{With this definition, when $ \theta =0 $, $ \phi $ corresponds to the angle between the origin and the position of the halo center.}
\begin{eqnarray}
	\vec{h} - \vec{\theta} = r (\cos\phi, \sin\phi)^\top, 
\end{eqnarray} 
and write
\begin{equation}
	\pdv{\theta_i} \Psi_{\rm halo} = \pdv{r}{\theta_i}\pdv{r}\Psi_{\rm halo}(r) = \frac{\theta_i - h_i}{r} \Psi'_{\rm halo}(r).
\end{equation}
In complex notation, we thus have
\begin{equation}
	\grad_{\rm c} \Psi_{\rm halo} = -\e^{\iu\phi} \Psi'_{\rm halo}(r).
\end{equation}
Notice the minus sign, which is there due to our choice of coordinates; it is easy to see that
\begin{equation}
	\pdv{\phi}{\theta_1} = \frac{\sin\phi}{r} \ , \ \pdv{\phi}{\theta_2} = -\frac{\cos\phi}{r},
\end{equation}
which has opposite sign with respect to the usual choice of cylindrical coordinate.
We can thus write 
\begin{equation}
	\grad_{\rm c} = -\e^{\iu\phi} \qty( \pdv{r}  + \frac{\iu}{r} \pdv{\phi}),
\end{equation} 
so that
\begin{align}
		&\Delta\kappa = \frac{1}{2} \qty( \Psi''_{\rm halo} + \frac{\Psi'_{\rm halo}}{r})  , \\ 
		&\Delta\gamma = \frac{\e^{2\iu\phi}}{2} \qty( \Psi''_{\rm halo} - \frac{\Psi'_{\rm halo}}{r}) , \\ 
		&G = -\frac{\e^{3\iu\phi}}{2} \qty( \Psi'''_{\rm halo} - 3\frac{\Psi''_{\rm halo}}{r} + 3\frac{\Psi'_{\rm halo}}{r^2}) , \\
		&F = -\frac{\e^{\iu\phi}}{2} \qty( \Psi'''_{\rm halo} + \frac{\Psi''_{\rm halo}}{r} - \frac{\Psi'_{\rm halo}}{r^2}) = -\e^{\iu\phi} \dv{\Delta\kappa}{r} .
\end{align}


We now specify to the NFW model. Here we discuss the spherical model without ellipticity, commenting on ellipticity App.~\ref{app:ellitp_nfw}. 
Using a coordinate system centered on the group, the NFW lensing potential is ~\cite{Bartelmann:1996hq}
\begin{align}
	&\Psi_{\rm NFW} (\theta) = 2\tilde\kappa\, \theta^2_{\rm s} \qty(\log^2\qty(\frac{\theta}{2\theta_{\rm s}}) - \rm{arccosh}^2\qty(\frac{\theta_{\rm s}}{\theta})  ) , \\ 
	&\theta_{\rm s} :=  \frac{R_{\rm s}}{d_{\rm l}} , \ \tilde\kappa := \frac{\rho_0 R_{\rm s}}{\Sigma_{\rm c}} , \ \Sigma_{\rm c} = \frac{d_{\rm s}}{4\pi Gd_{\rm l} d_{\rm ls}  } .\label{eq:ktildedef}
\end{align}

Defining $x=\theta_{\rm s}/h$, we have: 
\begin{align}
	&\Delta\beta_0 =  4h\,\tilde{\kappa}\,x^2 \qty(\log2x-\frac{\mathrm{arccosh}\,x}{\sqrt{1-1/x^2}})  \ , \\
	&\Delta\kappa = \tilde{\kappa}\frac{2x^2}{1 - x^2} \qty( 1 - \frac{ \mathrm{arccosh}\,x }{\sqrt{1-1/x^2}} ) \ ,\label{eq:kappanfw} \\ 
	&\Delta\gamma_0 = \tilde{\kappa}\frac{2x^2}{1 - x^2}  \Bigg( 1 + \frac{(2x^2-3) \mathrm{arccosh}\,x }{\sqrt{1-1/x^2}} \no\\ &+ 2\qty( 1 - x^2 )\log(2x)   \Bigg) , \label{eq:gamma_0} \\ 
	&F_0 = \frac{\tilde{\kappa}}{h}\frac{2x^2}{(1 - x^2)^2}  \qty( 2 + x^2 - \frac{3 \mathrm{arccosh}\,x }{\sqrt{1-1/x^2}}   ) , \label{eq:F_0} \\
	&G_0 = \frac{\tilde{\kappa}}{h}\frac{2x^2}{(1-x^2)^2} \Bigg(-\frac{15 - 20x^2 
	+8x^4}{\sqrt{1-1/x^2}}\mathrm{arccosh}\,x\nonumber \\ &+6-3x^2+8\left(1-x^2\right)^2\log(2x) \Bigg) . \label{eq:G_0}
\end{align}
Note that applying these expressions for $x<1$ requires using the logarithmic definition of $ \rm{arccosh}(z) = \ln(z + \sqrt{z^2-1}) $ and allowing complex $z$.

\begin{figure}
    \centering
    \includegraphics[scale=0.4]{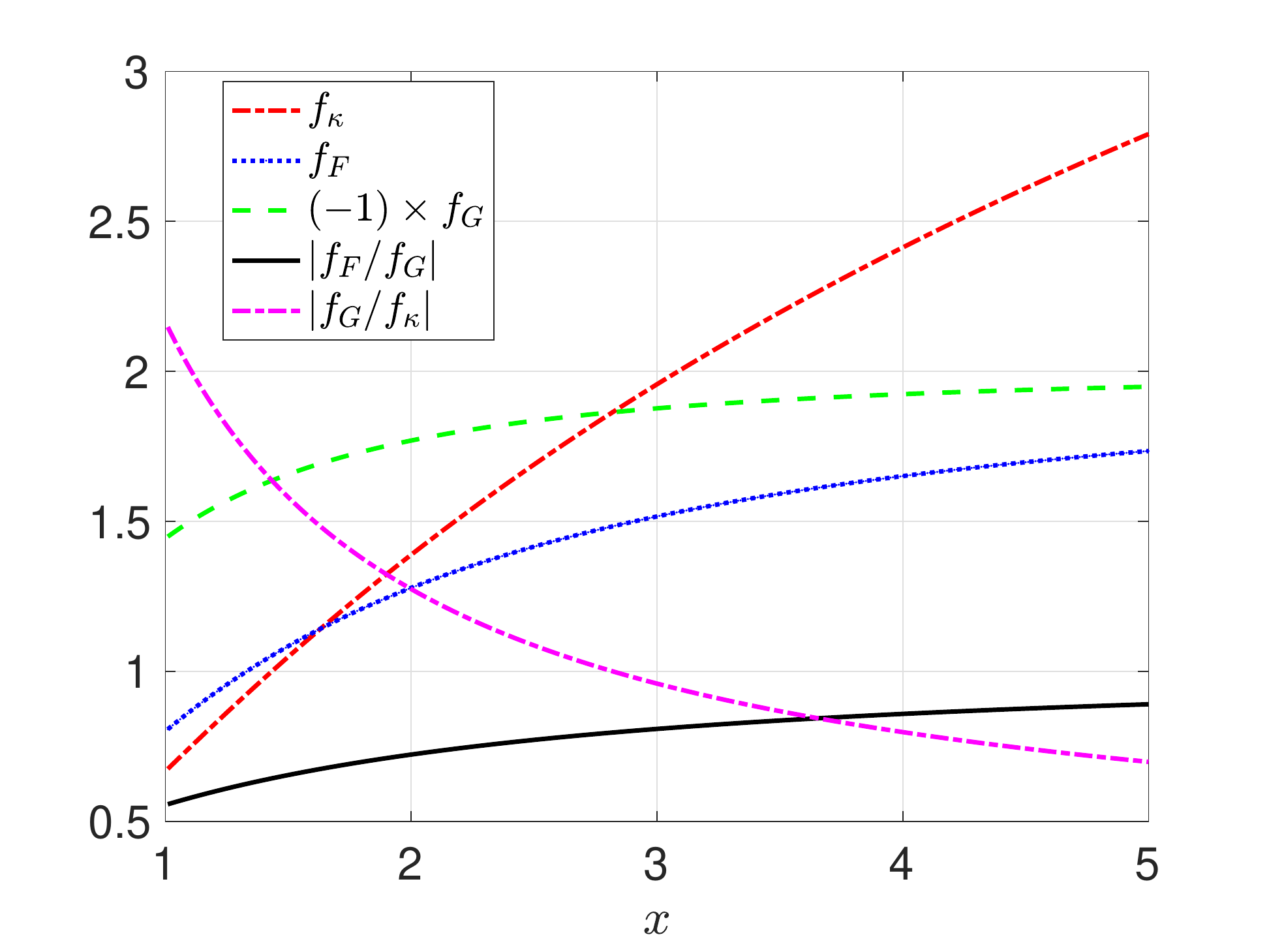}
    \caption{$x$-dependence of lensing expansion terms.}
    \label{fig:degkap}
\end{figure}

It is useful to define $f_{\kappa}(x),\,f_G(x),\,f_F(x)$ such that $\Delta\kappa=\tilde\kappa\,f_{\kappa}(x)$, $F_0=\frac{\tilde\kappa}{h}\,f_{F}(x)$, $G_0=\frac{\tilde\kappa}{h}\,f_{G}(x)$. These functions are shown in Fig.~\ref{fig:degkap}.

We can use these expressions to analyze the modeling constraints obtained from imaging data. As we already remarked in Sec.~\ref{s:group}, of the expansion terms, $\Delta\beta_0$ is exactly degenerate with the unknown source position; $\Delta\kappa$ is degenerate with external convergence, and can be absorbed by the MSD (changing the inference of $H_0$); $\Delta\gamma_0$ is degenerate with external shear; and so, only $F_0$ and $G_0$ can produce useful modeling constraints. However, since the axisymmetric NFW model contains three free parameters $\tilde\kappa,\,h,\,x=\theta_{\rm s}/h$, even a perfect measurement of the $F_0$ and $G_0$ terms still leaves a degeneracy.

\subsection{MSD}
Consider the usual MSD transformation, induced by a parameter $\lambda$:
\be\Delta\kappa&\to&\Delta\kappa_\lambda=\lambda\Delta\kappa+1-\lambda,\\
F_0&\to&F_{0,\lambda}=\lambda F_0,\;\;\;G_0\,\to\,G_{0,\lambda}=\lambda G_0\ee
(with matching transformation on the other terms, that are not relevant here). The combinations $F_{\rm eff}$ and $G_{\rm eff}$ of Eq.~\eqref{eq:reducedFlexion} 
are MSD-invariant, and are the quantities that constrain the model parameters. Suppose then that we are given precise determination of $F_{\rm eff}$ and $G_{\rm eff}$ from the imaging. In this case, the ratio $F_{\rm eff}/G_{\rm eff}$ determines $x$ . Having fixed $x$, there remains a degeneracy in $\Delta\kappa$, that we can express as
\be\label{eq:degkap} \Delta\kappa&=&\frac{h\,G_{\rm eff}\,f_\kappa(x)}{f_G(x)+h\,G_{\rm eff}\,f_\kappa(x)}.\ee
In Eq.~(\ref{eq:degkap}), we think of $G_{\rm eff}$ and $x$ as fixed by the imaging data, while the model parameter $h$ is free to vary (up to possible external priors, discussed in the main text). 

For $|h\,G_{\rm eff}|\ll1$, the parameter regime of most interest for us, we have $\Delta\kappa\approx h\,G_{\rm eff}\,f_\kappa(x)/f_G(x)$. Namely, an imaging determination of the flexion terms $G_{\rm eff}$ and $F_{\rm eff}$ cannot determine $\Delta\kappa$, which is almost directly degenerate with a change in the model parameter $h$ while holding $x=\theta_{\rm s}/h$ and $\tilde\kappa/h$ fixed.  

In terms of the original NFW model parameters, the $\Delta\kappa$ degeneracy maps to adjusting $R_{\rm s}\propto h$ while holding $\rho_0$ fixed. External priors are needed to break this degeneracy. A prior on $h$, the cluster center of mass position, obviously ameliorates it. A prior on the group velocity dispersion also ameliorates it, since $\sigma_{\rm los}^2\propto \rho_0R_{\rm s}^2$. The important point is that the external priors are crucial: without them, the MSD associated with an NFW group is not broken by the imaging data, even with explicit modeling of the group. 

\subsection{$G$-term degeneracy}
The threefold $G$-term degeneracy can also be clarified using the analytic expansion. To this end, it is useful to replace the model parameter $\tilde\kappa$ by $\Delta\kappa$. As we have seen in Sec.~\ref{s:Bias}, an MCMC trapped near a displaced likelihood minimum will attempt to reduce $ |F_{\rm eff}| $ while keeping $ G_{\rm eff} $ fixed. This amounts to flowing towards $x$ that minimizes $\left|F_{\rm eff}/G_{\rm eff}\right|=|f_F(x)/f_G(x)|$, while adjusting $\Delta\kappa$ and $h$ so as to keep $G_{\rm eff}=(f_G/f_\kappa)(\Delta\kappa/h)/(1-\Delta\kappa)$=Const. As can be seen from Fig.~\ref{fig:degkap}, minimizing $|f_F(x)/f_G(x)|$ pulls the fit towards smaller $x$. In turn, this pulls $|f_G(x)/f_\kappa(x)|$ to a larger value, meaning that in order to compensate and keep $G_{\rm eff}$ constant, the combination $\left(\Delta\kappa/h\right)/\left(1-\Delta\kappa\right)$ is pulled to a smaller value. For $\Delta\kappa\ll1$, this tends to make the fit pull towards a model in which $\Delta\kappa$ is smaller than its truth value, causing a positive upwards bias in $H_0$.

\subsection{Time delays}
Time delays between image A and B can be written as (see e.g. \cite{Schneider:1992,Bartelmann:1999yn})
\begin{align}
	\begin{aligned}
	\Delta t_{\rm AB} &= D_{\rm dt} \qty( \frac{\theta^2_{\rm A}}{2} - \vec\beta \cdot \vec\theta_{\rm A} - \Psi(\vec\theta_{\rm A}) - (\mathrm{A} \leftrightarrow \mathrm{B}) ) \\
	&=: D_{\rm dt} (\tau_{\rm A} - \tau_{\rm B}) \ , \ D_{\rm dt} = (1+z_{\rm l}) \frac{d_{\rm l}d_{\rm s}}{d_{\rm ls}} .	
	\end{aligned}
\end{align}
One can find the time delay from the complex lens equation, by noticing
\begin{equation}
	\grad_{\rm c} = 2\pdv{\theta^*} \implies \tau = \frac{1}{2} \int \dd{\theta^*} \alpha + f(\theta) \ ,
\end{equation}
where the integral is a definite integral. Notice that the integral misses a possible function of $ \theta $ alone; this function can be recovered by imposing $ \tau = \tau^* $.

Focusing on $ \tau_{\rm A} $, in complex notation we can write
\begin{align}
	\begin{aligned}
		\tau_{\rm A} &= \frac{\theta_{\rm A}\theta^*_{\rm A}}{2} -\frac{1}{2} \qty[ (\beta + \grad_{\rm c}\Psi_{\rm nfw}) \theta^*_{\rm A} + \mathrm{c.c.} ] - \Psi_{\rm lens}(\vec\theta_{\rm A}) \\
		&- \Psi_{\rm halo}(0,0) - \frac{1}{2} \kappa \theta_{\rm A}\theta^*_{\rm A} - \frac{1}{4} \qty[(\gamma + \gamma^{\rm ext}){\theta^*_{\rm A}}^2 + \mathrm{c.c.}]\\
		&-\frac{1}{24}\qty(3F{\theta^*_{\rm A}}^2 \theta_{\rm A} + G{\theta^*_{\rm A}}^3 + \mathrm{c.c.} ) .
	\end{aligned}
\end{align}
In real notation,
\begin{align}
	\begin{aligned}
		\tau_{\rm A} &= \frac{\theta^2_{\rm A}}{2} - (\beta_i + \partial_i\Psi_{\rm nfw}) \theta_i^{\rm A} - \Psi_{\rm lens}(\vec\theta_{\rm A}) - \Psi_{\rm halo}(0,0) \\
		&- \frac{1}{2} \kappa \theta^2_{\rm A} - \frac{1}{2} (\gamma_1 + \gamma^{\rm ext}_1)({\theta^{\rm A}}^2_1
		-{\theta^{\rm A}}^2_2)-  (\gamma_2 + \gamma^{\rm ext}_2)\theta^{\rm A}_1 \theta^{\rm A}_2 \\
		& -\frac{1}{12}\Big((3F_1 + G_1){\theta^{\rm A}}^3_1 + (3F_2 - G_2){\theta^{\rm A}}^3_2  \\
		&+ 3(F_1 - G_1)\theta^{\rm A}_1{\theta^{\rm A}}^2_2 + 3(F_2 + G_2)\theta^{\rm A}_2{\theta^{\rm A}}^2_1\Big).
	\end{aligned}
\end{align}

\subsection{Effect of ellipticity} \label{app:ellitp_nfw}
We can implement ellipticity in the NFW profile by using pseudo-elliptical NFW lens models~\cite{Golse:2001ar,Dumet-Montoya:2012ynz}, which is a reliable approximation as long as the ellipticity is not too large. 
This approximation, implemented in \texttt{lenstronomy}, uses the spherical NFW lensing potential, computed in elliptical coordinates,
\beq
\Psi_\epsilon(\vec{\theta}) = \Psi(\vec{\theta}_\epsilon) \ , \  \vec{\theta}_\epsilon = 
\begin{pmatrix}
	\cos\varphi \sqrt{1-\epsilon} & 	\sin\varphi \sqrt{1-\epsilon}\\
		-\sin\varphi \sqrt{1+\epsilon} & 	\cos\varphi \sqrt{1+\epsilon}\\
\end{pmatrix} \vec{\theta} ,
\eeq
where 
\beq
\epsilon = \frac{2e}{1+e^2} , \ e = \sqrt{e_1^2 + e_2^2} , \ \varphi = \frac{1}{2} \arctan \frac{e_1}{e_2} .
\eeq
Notice that under rotations of the coordinate system, $ \epsilon $ is invariant but $ \varphi $ changes.
The expansion of the potential $ \Psi_{\mathrm{halo},\epsilon} $ follows Eq.~\eqref{eq:expansion}; however, since the potential is not axisymmetric, it is not possible to factor out the $ \phi $ dependence on $ \Delta\gamma_\epsilon $, $ G_\epsilon $, $ F_\epsilon $ as we did in Eq.~\eqref{eq:sph_sym_expansion}. Nevertheless, it is possible to express these quantities as functions of $ \Delta\gamma_0 $, $ F_0 $ and $ G_0 $. 

Notice that we can write (here, $ \vec{\theta} = (x , y)^\top $) 
\beq
\grad_{\rm c} = \e^{\iu\varphi} \qty(\sqrt{1-\epsilon} \pdv{x_\epsilon} + \iu\sqrt{1+\epsilon} \pdv{y_\epsilon} ).  
\eeq
As an example, consider
\begin{align}
\gamma_\epsilon &= \frac{1}{2} \grad_{\rm c}\grad_{\rm c}\Psi_\epsilon = \frac{\e^{2\iu\varphi}}{2} \Big( (1-\epsilon)\pdv{^2}{x^2_\epsilon} - (1+\epsilon)\pdv{^2}{y^2_\epsilon} \no\\
&+ 2\iu \sqrt{1-\epsilon^2} \pdv{x_\epsilon}\pdv{y_\epsilon}\Big)\Psi_\epsilon\no\\
&= \e^{2\iu\varphi}(\gamma_1(\vec{\theta}_\epsilon)  + \iu \sqrt{1-\epsilon^2}\gamma_2(\vec{\theta}_\epsilon) - \epsilon\kappa(\vec{\theta}_\epsilon)) ,
\end{align}
where the quantities without $ _\epsilon $ subscript refer to the spherical functions computed at the elliptic coordinate.
By defining the displacement vector in elliptical coordinates in the complex notation, $ \vec{h} \leftrightarrow  h_\epsilon\e^{\iu\phi_e} $, we can express 
\beq
\gamma = \gamma_0(h_\epsilon(h,\phi))\e^{\iu \phi_\epsilon(\phi)}  .
\eeq
For the NFW example, $ \gamma_0 $ is the expression in Eq.~\eqref{eq:gamma_0}.
For $ \kappa $, one has
\begin{equation}
	\kappa_\epsilon = \kappa(h_\epsilon(h,\phi)) -\epsilon\gamma_1(h_\epsilon(h,\phi))   .
\end{equation}
A similar reasoning follows for $ F_\epsilon $, $ G_\epsilon $. 
As one would expect, ellipticity modifies the effective flexion and convergence by correction terms of order $ \epsilon $. Our numerical MCMC analysis includes the full effect, and moderate or small ellipticity has only a minor effect on the results.

\section{Kinematics constraints for host group}\label{app:kin}
Here we review constraints on a host group, obtainable by measurements of a sample of galaxy members. We only consider spherical systems with an isotropic velocity distribution. We start with a PL mass and tracer galaxy distribution, and go on to consider the NFW profile. 
For clarity, our discussion repeats a number of statements from the main text, completing these with details and derivations.

\subsection{Power law}
We start with the PL profile, $\rho(r)=\rho_0\left(r/R_0\right)^{-\gamma}$. The surface density of this profile is
\be\Sigma&=&
\frac{\sqrt{\pi}\Gamma\left(\frac{\gamma-1}{2}\right)\rho_0R_0}{\Gamma\left(\frac{\gamma}{2}\right)}\left(\frac{d_{\rm l}\theta}{R_0}\right)^{1-\gamma},\ee
resulting with the deflection angle and convergence,
\be\alpha&=&\left(\frac{\theta}{\theta_{\rm E}}\right)^{1-\gamma}\theta,\,\,\,\,\kappa\,=\,\frac{3-\gamma}{2}\left(\frac{\theta}{\theta_{\rm E}}\right)^{1-\gamma},
\ee
where
\be\theta_{\rm E}&=&\left[\frac{2\sqrt{\pi}\Gamma\left(\frac{\gamma-1}{2}\right)}{(3-\gamma)\Gamma\left(\frac{\gamma}{2}\right)}\right]^{\frac{1}{\gamma-1}}\frac{R_0}{d_{\rm l}}\left(\frac{\rho_0R_0}{\Sigma_c}\right)^{\frac{1}{\gamma-1}}.\ee

The circular velocity is
\be v_{\rm circ}^2&=&\frac{GM(r)}{r}\,=\,\frac{4\pi G\rho_0R_0^2}{3-\gamma}\left(\frac{d_{\rm l}\theta}{R_0}\right)^{2-\gamma}.\ee
The circular velocity is not directly measurable. What is measurable is the LOSVD~\cite{2008gady.book.....B},
\be\sigma_{\rm los}^2&=&\frac{2G}{s_*(r)}\int_1^\infty \frac{dy\,y}{\sqrt{y^2-1}}\,\int_{y}^\infty\frac{dx}{x^2}n_*(xr)M(xr)\label{eq:losvd0}\\
&=&\frac{8\pi G\rho_0R_0^3r^{3-\gamma}}{(3-\gamma)s_*(r)}\int_1^\infty \frac{dy\,y}{\sqrt{y^2-1}}\,\int_{y}^\infty dx\,n_*(xr)\,x^{1-\gamma},\no\\&&\label{eq:losvdPL1}\ee
where $n_*$ and $s_*$ are the galaxy number density and surface density, respectively. 
Eq.~(\ref{eq:losvd0}) holds for an arbitrary isotropic profile, while  Eq.~(\ref{eq:losvdPL1}) applies to the PL. 
If the galaxy number density is distributed similarly to the mass density in the group, $n_*(r)\propto (r/R_0)^{-\gamma}$, we find (valid for $\frac{3}{2}<\gamma<\frac{5}{2}$):
\be\label{eq:losvdPL}\sigma_{\rm los}^2&=&\frac{\Gamma^2\left(\frac{\gamma}{2}\right)\Gamma\left(\gamma-\frac{3}{2}\right)}{4\sqrt{\pi}\,\Gamma^2\left(\frac{\gamma-1}{2}\right)\Gamma\left(\gamma\right)} \frac{d_{\rm s}}{d_{\rm ls}}\theta_{\rm E}\left(\frac{\theta}{\theta_{\rm E}}\right)^{2-\gamma}.
\ee
For the SIS ($\gamma=2$),  
$\theta_{\rm E}
=8\pi^2\frac{d_{\rm ls}}{d_{\rm s}  }G\rho_0R_0^2$ and  $\sigma_{\rm los}^2
=\frac{d_{\rm s}}{d_{\rm ls}  }\frac{\theta_{\rm E}}{4\pi}=\frac{v_{\rm circ}^2}{2}$. 

The relation $\sigma_{\rm los}^2=\frac{d_{\rm s}}{d_{\rm ls}  }\frac{\theta_{\rm E}}{4\pi}$ is sometimes adopted by lensing analyses. However, the expressions for $\sigma_{\rm los}^2$  depend on a series of simplifying assumptions, including: PL mass distribution; same PL galaxy number distribution; virial state (following Jeans equation); spherical symmetry; isotropic velocity distribution; and to satisfy $\sigma_{\rm los}^2=\frac{d_{\rm s}}{d_{\rm ls}  }\frac{\theta_{\rm E}}{4\pi}$, the specific case of the SIS. We do not expect those assumptions to hold precisely for individual galaxy groups. Lensing analyses would be prudent to assign a larger systematic uncertainty than the ``bare" observational uncertainty on $\sigma^2_{\rm los}$.

\begin{figure}
    \centering
    \includegraphics[scale=0.35]{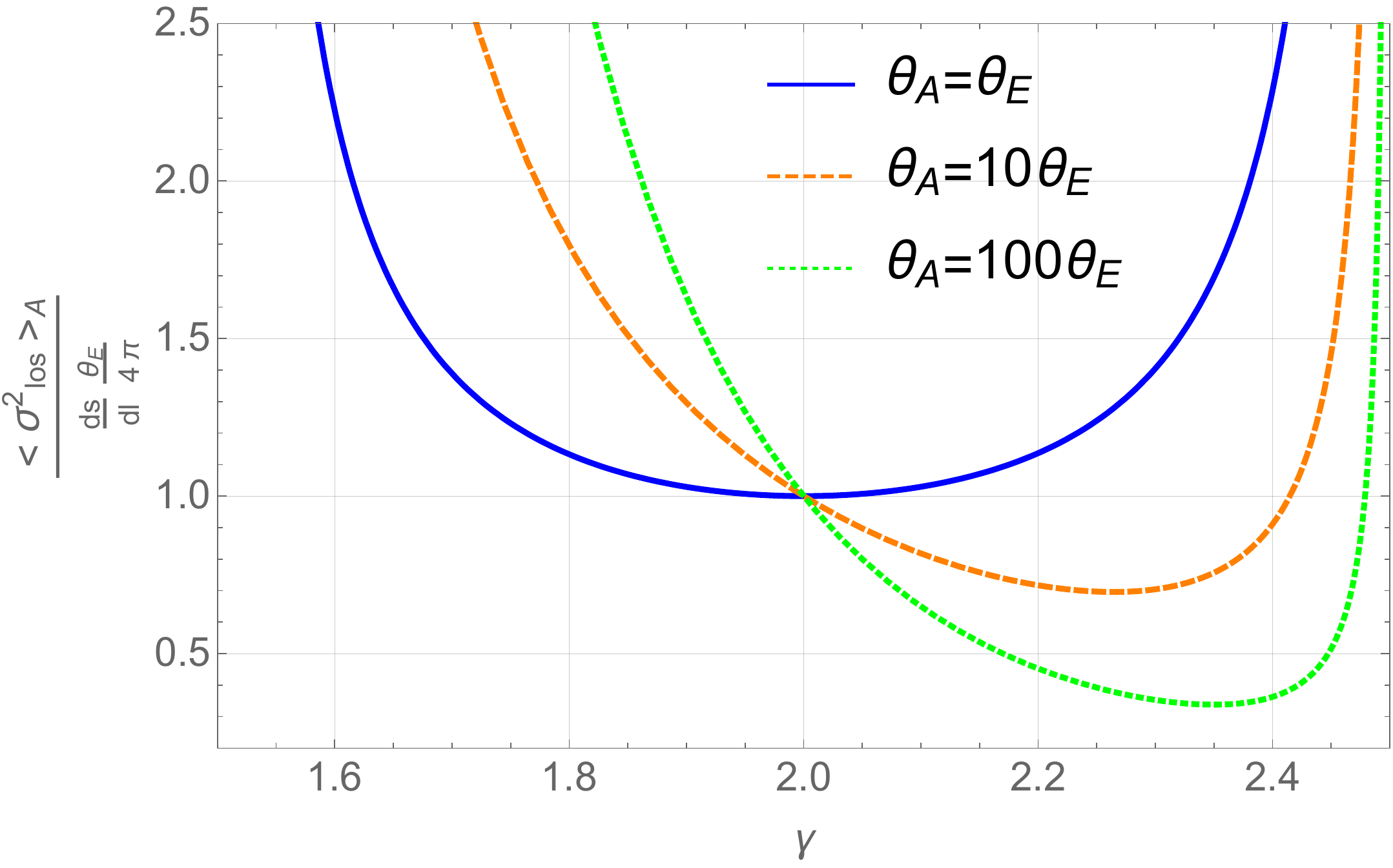}
    \caption{$\gamma$- and $\theta_A$- dependence of aperture-averaged LOSVD.}
    \label{fig:losvdPL}
\end{figure}

To explore one of those effects, consider varying $\gamma$ in the relation between $\sigma_{\rm los}^2$ and $\theta_{\rm E}$. For non-SIS systems, $\sigma_{\rm los}^2(\theta)$ depends on $\theta$, and a relevant observable is the brightness-weighted average of $\sigma_{\rm los}^2$ in some aperture $\theta_A$. We can compare such averaged LOSVD to the SIS relation:
\be\frac{\langle\sigma_{\rm los}^2\rangle_A}{\frac{d_{\rm s}}{d_{\rm ls}  }\frac{\theta_{\rm E}}{4\pi}}&=&\frac{1}{\frac{d_{\rm s}}{d_{\rm ls}  }\frac{\theta_{\rm E}}{4\pi}}\frac{\int_0^{\theta_A} d\theta\theta s_*(\theta)\sigma^2_{\rm los}(\theta)}{\int_0^{\theta_A} d\theta\theta s_*(\theta)}\label{eq:losvdPL0}\\
&=&\frac{3-\gamma}{5-2\gamma}\frac{\sqrt{\pi}\Gamma^2\left(\frac{\gamma}{2}\right)\Gamma\left(\gamma-\frac{3}{2}\right)}{\Gamma^2\left(\frac{\gamma-1}{2}\right)\Gamma\left(\gamma\right)} 
\left(\frac{\theta_{\rm A}}{\theta_{\rm E}}\right)^{2-\gamma}.\;\;\;\;\;\label{eq:losvdPL}\ee
We illustrate Eq.~(\ref{eq:losvdPL}) in Fig.~\ref{fig:losvdPL}.

\subsection{NFW} \label{subsec:NFW_kinematics}
\begin{figure}
    \centering
    \includegraphics[scale=0.475]{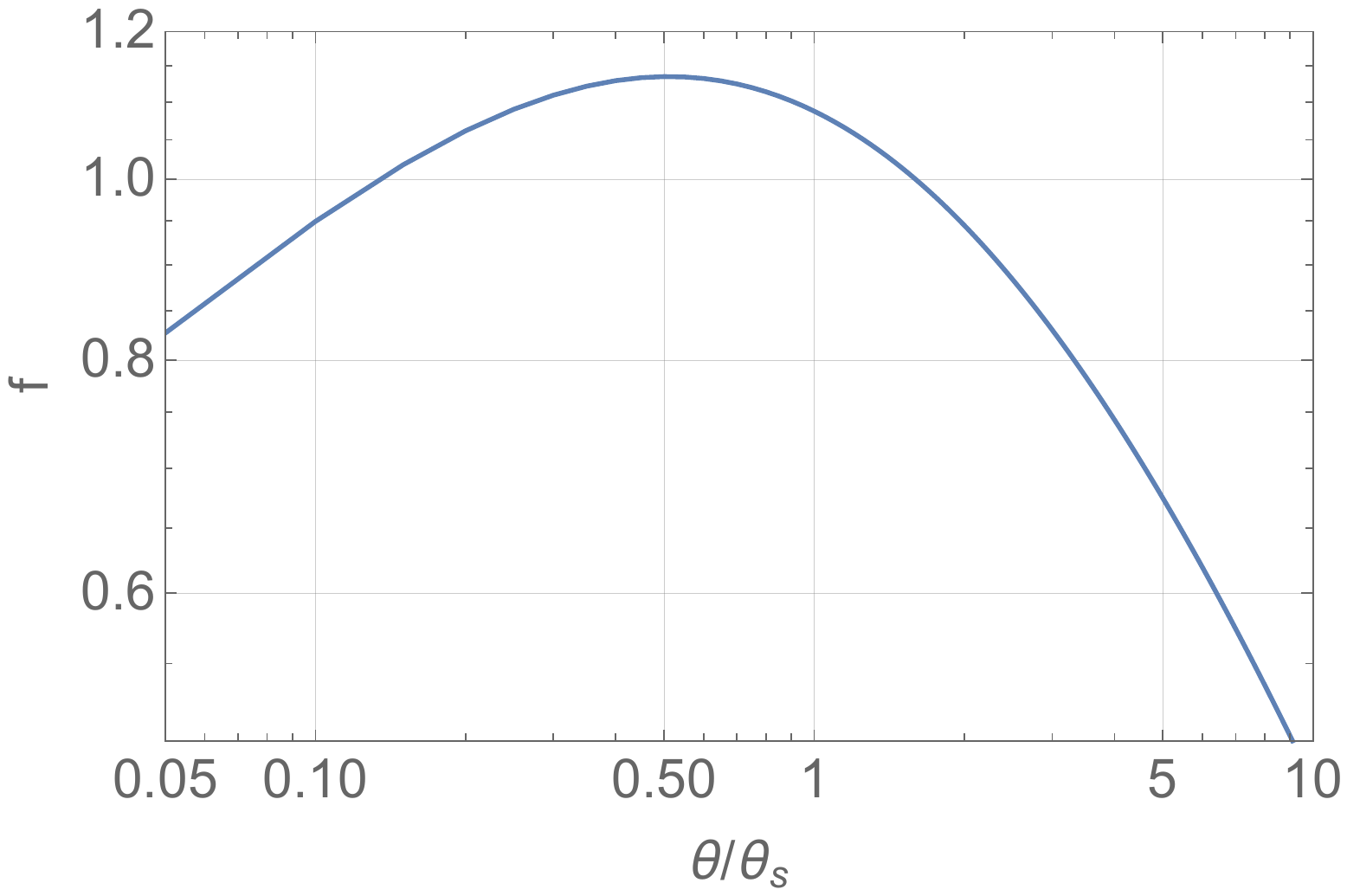}
    \caption{The LOSVD $\theta$-dependence for the NFW profile.}
    \label{fig:fNFW}
\end{figure}

For the NFW profile, assuming again that the tracer galaxy density follows the mass density, Eq.~(\ref{eq:losvd0}) can be written as:
\be\label{eq:losvdNFW}\sigma_{\rm los}^2(\theta)&=&G\rho_0R_{\rm s}^2\,f\left(\frac{\theta}{\theta_{\rm s}}\right),\\
f(a)&=&\frac{8\pi a\int_1^\infty \frac{dy y}{\sqrt{y^2-1}}
\int_{ya}^\infty\frac{dx}{x^3\left(1+x\right)^2}\left(\ln\left(1+x\right)-\frac{x}{1+x}\right)}{2\int_0^\infty \frac{dz}{\sqrt{a^2+z^2}(1+\sqrt{a^2+z^2})^2}}.\no\\&&
\ee
Some of these integrals can be done in closed form, but that is not particularly illuminating. 
The function $f(\theta/\theta_{\rm s})$ is shown in Fig.~\ref{fig:fNFW}. 

\begin{figure}
    \centering
    \includegraphics[scale=0.425]{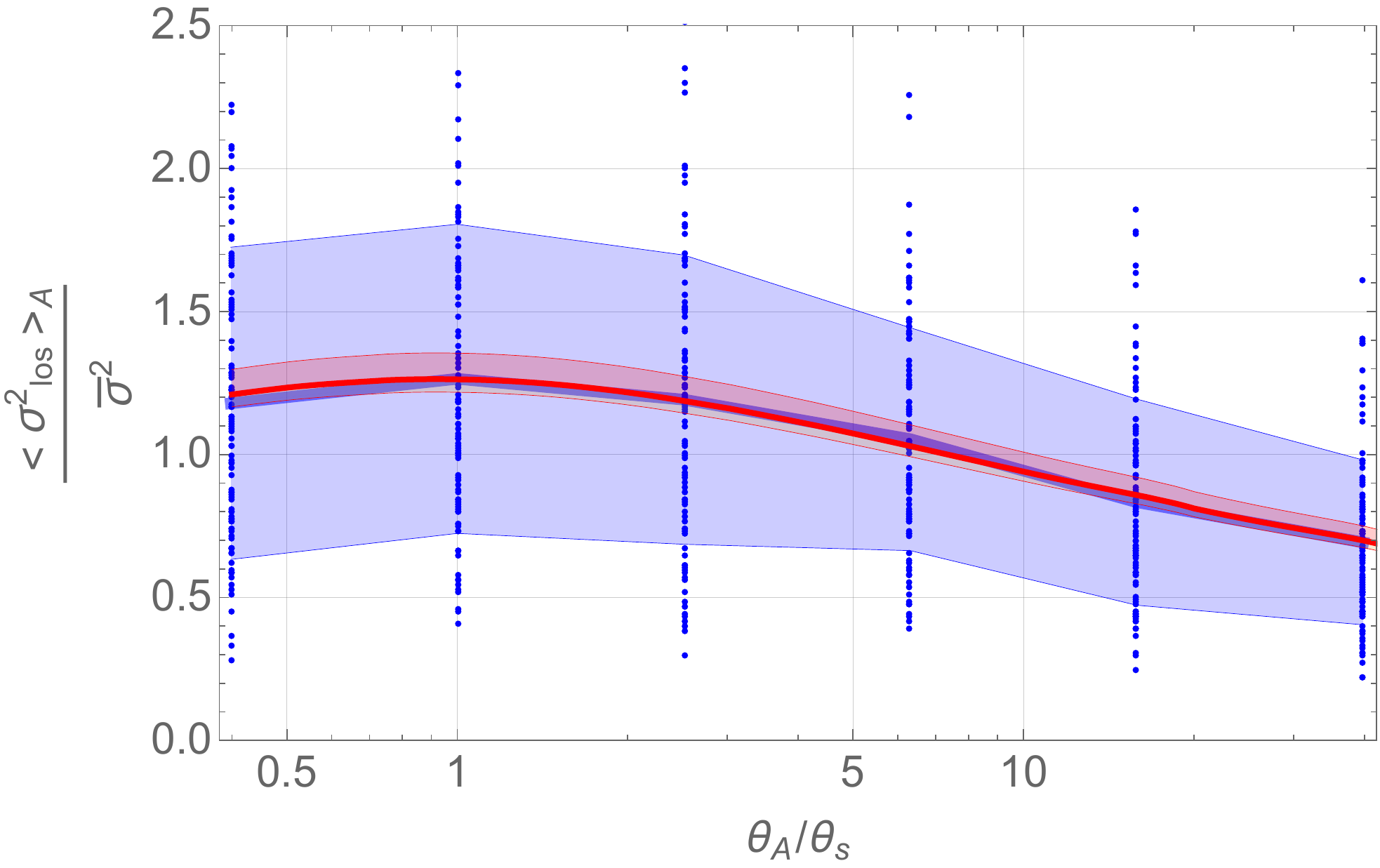}
    \caption{Aperture-averaged LOSVD for NFW, compared to the quantity $\bar\sigma^2=GM_{\rm vir}/(3R_{\rm vir})$. Red line: using Eq.~(\ref{eq:losvdNFW}) in Eq.~(\ref{eq:losvdPL0}), replacing $\frac{d_{\rm s}}{d_{\rm ls}  }\frac{\theta_{\rm E}}{4\pi}$ by $\bar\sigma^2$, with $c_{\rm vir}=3.5$. Red shaded band: varying $c_{\rm vir}\to3.5\times10^{\pm0.14}$. Dots: direct calculation of $\sigma_{\rm los}^2$ from mock groups, assuming 13 tracer galaxies per realization (see Sec.~\ref{ss:sampvar}). Blue line (somewhat hidden under the red) and blue shaded band: mean and $1\sigma$ region for the mocks.}
    \label{fig:losvdNFWtoGMR}
\end{figure}

Assuming the NFW model, a kinematics prior should use Eq.~(\ref{eq:losvdNFW}) (aperture-averaged as needed) to translate a LOSVD measurement into a constraint on the combination of $\rho_0$ and $R_{\rm s}$ appearing in the equation. However, analyses in the literature took a somewhat different route. 
Ref.~\cite{Chen:2019ejq} (after~\cite{Koranyi:1999jz,Wong:2010xk}) considered the following expression as a proxy for the LOSVD of the NFW profile:
\be \label{eq:sigma}
\bar\sigma^2&=&\frac{GM_{\rm vir}}{3R_{\rm vir}},\ee
where $R_{\rm vir}=c_{\rm vir}\,R_s$, $M_{\rm vir}=M(R_{\rm vir})$, and $c_{\rm vir}$ is the  NFW concentration parameter, for which it is possible to extract a theoretical prediction from N-body simulations~\cite{Maccio:2008pcd}. Namely, instead of using the physical relation, Eq.~(\ref{eq:losvdNFW}), to convert the measured LOSVD into a constraint on the group halo model, Ref.~\cite{Chen:2019ejq} made the identification $\sigma_{\rm los}^2\to\bar\sigma^2$, and then used Eq.~(\ref{eq:sigma}) to constrain $M_{\rm vir}$ and $R_{\rm vir}$. 
The dependence of the RHS of Eq.~(\ref{eq:sigma}) on $c_{\rm vir}$ can be clarified by noting that
\be\label{eq:r0rssbar}\bar\sigma^2&=&G\rho_0R_s^2\frac{4\pi}{3c_{\rm vir}}\left(\ln(1+c_{\rm vir})-\frac{c_{\rm vir}}{1+c_{\rm vir}}\right).\ee
The dependence is not strong: the factor $\frac{4\pi}{3c_{\rm vir}}\left(\ln(1+c_{\rm vir})-\frac{c_{\rm vir}}{1+c_{\rm vir}}\right)$ is equal to $\left\{0.89,0.85,0.80,0.62\right\}$ when varying $c_{\rm vir}=\left\{3,4,5,10\right\}$, respectively. However, it is noteworthy that this dependence is not physical, but rather introduced artificially. 
The red solid line in Fig.~\ref{fig:losvdNFWtoGMR} compares the physical aperture-weighted LOSVD to the quantity $\bar\sigma^2$, using $c_{\rm vir}=3.5$, relevant for low-redshift massive galaxies. 

The meaning of Fig.~\ref{fig:losvdNFWtoGMR} is that using Eq.~(\ref{eq:sigma}) and identifying $\sigma^2_{\rm los}\to\bar\sigma^2$ can introduce a systematic error. The magnitude of the error depends on the aperture and on the value adopted for $c_{\rm vir}$, but can amount to $\sim20\%$.

\subsection{Kinematical and cosmological priors}\label{ss:kinprior}
Here we describe our attempt to define a kinematics+cosmology prior, following a similar procedure as used in~\cite{Chen:2019ejq} for PG1115+080.

The first step taken by~\cite{Chen:2019ejq} is to identify $\sigma_{\rm los}\to\bar\sigma$. 

The next step is to invoke theoretical expectations based on N-body simulations. Following App. A of~\cite{Wong:2010xk}, the virial mass and the virial radius of the halo are related to a characteristic redshift-dependent overdensity:
\be \label{eq:M_kin}
M_{\rm vir} &=& \frac{4\pi}{3} R^3_{\rm vir} \Delta_{\rm c}(z) \rho_{\rm c}(z) ,
\ee
where 
$ \Delta_{\rm c}(z) = 178\qty(\rho_{\rm m}(z)/\rho_{\rm c}(z))^{0.45}$
is the expected halo overdensity at the virial radius and $ \rho_{\rm c}(z) $ and $ \rho_{\rm m}(z)$ are the cosmological critical density and matter density at redshift $ z $~\cite{Eke:2000av}. 
Combining Eqs.~(\ref{eq:M_kin}) and~(\ref{eq:sigma}) turns $ \bar\sigma $ into separate priors on $ R_{\rm vir} $ and $M_{\rm vir}  $:
\be \label{eq:Rvir}R_{\rm vir}&=&\frac{3\bar\sigma}{2\sqrt{\pi G \Delta_{\rm c} \rho_{\rm c}}},\\
\label{eq:Mvir}M_{\rm vir}&=&\frac{9\bar\sigma^3}{2\sqrt{\pi G^3\Delta_{\rm c}\rho_{\rm c}}}.
\ee

Next, Ref.~\cite{Chen:2019ejq} adds a prior on $c_{\rm vir}$, using the $ M_{\rm vir}-c_{\rm vir} $ relation from Ref.~\cite{Maccio:2008pcd}:
\begin{eqnarray} \label{eq:c}
\log_{10} c_{\rm vir} &\approx& 0.97 - 0.09\log_{10}\qty(\frac{M_{\rm vir}\,h}{10^{12} M_\odot }).
\end{eqnarray}
Treating $c_{\rm vir}$ as a function of $\bar\sigma$ via Eqs.~(\ref{eq:c}) and~(\ref{eq:Mvir}), one has now produced separate priors on the parameters $\rho_0$ and $R_{\rm s}$:
\be\label{eq:rho0}\rho_0&=&\frac{\Delta_{\rm c}\rho_{\rm c} c^3_{\rm vir}}{3}\qty(\ln(1+c_{\rm vir}) -\frac{c_{\rm vir}}{1+c_{\rm vir}})^{-1},\\
\label{eq:Rs}R_{\rm s}&=&\frac{R_{\rm vir}}{c_{\rm vir}}. \ee
The uncertainty can be estimated by combining the observational uncertainty on $\sigma_{\rm los}$ and the theoretical scatter on the $M_{\rm vir}-c_{\rm vir}$ relation; the latter is taken in Ref.~\cite{Chen:2019ejq} as $\pm0.14$ on the $\log_{10}c_{\rm vir}$ expression in Eq.~(\ref{eq:c}).

In practice, our implementation of these priors in the MCMC mock analysis is as follows: 
\begin{enumerate}
\item The  parameters that are directly sampled by the chain, defining the posterior likelihood space, are angular variables that are linearly proportional to $\rho_0$ and $R_{\rm s}$. In each step of the calculation, we translate the sampled value of $\rho_0$ into an expected value of $c_{\rm vir}$, using Eq.~\eqref{eq:rho0}. 
\item Having obtained $c_{\rm vir}$, we convert Eq.~(\ref{eq:c}) into an equation for $R_{\rm vir}$, replacing $M_{\rm vir}$ by $R_{\rm vir}$ using Eq.~(\ref{eq:M_kin}). Now, we implement the estimated theoretical scatter of $\pm0.14$ dex on $c_{\rm vir}$ in the $M_{\rm vir}-c_{\rm vir}$ relation, to define a range of acceptable values for $R_{\rm s}$ via $R_{\rm s}\subset R_{\rm vir}/(10^{+0.14},10^{-0.14})$. In the chain, we discard sampled values of $R_{\rm s}$ that fall outside of this range. This step therefore enforces a redshift-dependent correlation between $\rho_0$ and $R_{\rm s}$. Up to this point, we made no connection to kinematics. 
\item Finally we come to the kinematics. Using Eq.~(\ref{eq:r0rssbar}), we convert the model point represented by $\rho_0$ and $R_{\rm s}$, along with the central value of $c_{\rm vir}$, into a model prediction for $\bar\sigma$. This prediction is then compared to the measured value of $\sigma_{\rm los}$ using the nominal observational uncertainty to obtain a likelihood factor.
\end{enumerate}

\subsection{Sample variance}\label{ss:sampvar}
Sample variance is the dominant nominal source of uncertainty in $\sigma^2_{\rm los}$, given that typically only a handful of galaxies are measured as tracers of the group. Ref.~\cite{Wilson:2017apg} used bootstrap to estimate sample variance directly from the measured sample of galaxies; here, we complement this route by generating random sets of $N_{\rm g}$ galaxies tracing an NFW halo. 

The realizations are drawn from an equilibrium phase space distribution function $f(r,v)$. For a spherical halo with a statistically static distribution function, we have $f(r,v)=f(\varepsilon(r,v))$, where $\varepsilon=\psi(r)-\frac{v^2}{2}$, $\psi(r)=-\phi(r)$, and $\phi(r)$ is the halo Newtonian potential. $f(\varepsilon)$ is given by\footnote{See~\cite{2008gady.book.....B}, Ch.4.3.1.}
\be f(\varepsilon)&=&\frac{1}{\sqrt{8}\pi^2}\int_0^{\varepsilon}\frac{d\psi}{\sqrt{\varepsilon-\psi}}\frac{d^2\rho}{d\psi^2}.\ee
We calculate $f(\varepsilon)$ numerically, and use it to draw samples of $N_{\rm g}$ tracer galaxies that fall within projected aperture $\theta_{\rm A}$. The results of this exercise for $N_{\rm g}=13$ (as in PG1115+080) and for different apertures are shown as blue dots in Fig.~\ref{fig:losvdNFWtoGMR}. For each value of $\theta_{\rm A}/\theta_{\rm s}$ we generate 100 mock samples. For each sample we calculate $\sigma_{\rm los}^2$ directly as the variance of LOS velocity across the $N_{\rm g}$ galaxies. The mean and standard deviation of $\sigma_{\rm los}^2$ are shown by the thick blue line and shaded region.

The variance we find for $\sigma^2_{\rm los}$ in Fig.~\ref{fig:losvdNFWtoGMR} is roughly consistent with sample variance of a normal distribution: $\frac{\delta\sigma^2_{\rm los}}{\sigma^2_{\rm los}}\approx\sqrt{\frac{2}{N_g-1}}\approx0.4$ for $N_g=13$. For comparison with Ref.~\cite{Wilson:2016hcs}, we also calculate uncertainty estimates for $\sigma_{\rm los}^2$ using the bootstrap method. As a rule, the bootstrap method provides a slightly lower uncertainty estimate than the variance found with mock realizations, but the difference is small: whereas direct sample variance predicts $\Delta\sigma_{\rm los}^2/\sigma_{\rm los}^2\approx0.4$, bootstrap predicts $\Delta\sigma_{\rm los}^2/\sigma_{\rm los}^2\approx0.35$. Thus, we reproduce the sample variance-dominated uncertainty estimate of Ref.~\cite{Wilson:2017apg} for PG1115+080.

\section{Estimating the probability of an MCMC to fall into a displaced minimum}\label{ss:probfalse}
As noted in Sec.~\ref{ss:results}, at least for a wide group centroid prior (as derived in Ref.~\cite{Wilson:2017apg} for PG1115+080), trial and error with \texttt{emcee}~\cite{Foreman_Mackey:2013} suggests that the initial placement of the walkers is a key factor in deciding which likelihood minimum will attract the fit. We can (roughly) estimate the probability of falling into a wrong minimum by the probability for the walker placement to start off closer to a false minimum than to the truth one. Let us assume that the initial placement of the walkers is chosen to coincide with the prior's center (this seems like a natural choice). In this case, the probability to fall into a wrong minimum is roughly given by the probability of the group centroid prior to be nearer a false minimum than the truth one. 

We can estimate this probability using mock samples of tracer galaxies as in  Sec.~\ref{ss:sampvar}. Consider a sample of $N_{\rm g}$ galaxies, and choose a random member to be the ``primary lens". From the same sample, derive a group center prior as the center of mass of the members. In general, of course, the prior center does not coincide with the center of the halo used to generate the mock.\footnote{The statistical distribution of the mismatch, which goes to defining the prior width, can be estimated either by bootstrap, as done in~\cite{Wilson:2017apg} for the actual data, or by repeated mocks. We test both, and find them to be compatible.} Now, draw a line connecting the primary lens with the true halo center, and another line connecting the primary lens with the prior center: the prior center is closer to a false minimum if the smaller angle between these lines is larger than $60^{\rm o}$. 

The result of this calculation depends on the number of member galaxies $N_{\rm g}$, the group halo profile, and the analysis aperture. For $N_{\rm g}=13$ in the NFW model with an aperture of 10~$R_{\rm s}$ (around 3 virial radii), we find the wrong minimum probability to be $\sim10$\%. 

It should be clear that the simple estimate we described here ignores a range of possible selection effects, both natural (such as mass segregation, and the bias of more massive group members to become primary lenses) as well as analysis-specific (such as luminosity-dependent contamination and incompleteness). Such effects can probably modify our estimates at the $\mathcal{O}(1)$ level, but should not change the order of magnitude of the result.

\section{How common is the flexion degeneracy?}\label{app:d}
PG1115+080 is proof of concept that lensing set-ups like the one we analyzed are observationally relevant. But since PG1115+080 was selected specifically because of its massive, near-by group association, this set-up may be quite uncommon. Most of the lensed quasars analyzed for time-delay cosmography may either not be associated with LOS groups; or, if they are, may lie far from the group's projected center of mass, making flexion terms less important than for PG1115+080. A detailed analysis of the fraction of systems that may exhibit flexion degeneracy is beyond the scope of this work. Nevertheless we can make a crude estimate using the data from~\cite{Wilson:2017apg}, as follows. 

Tab.~1 in Ref.~\cite{Wilson:2017apg} lists reconstructed properties of LOS groups (including primary lens host where exists) of their 26 lens systems. For each LOS group, estimated values of $M_{\rm vir}$ and $R_{\rm vir}$ are quoted, along with the measured LOSVD, denoted there by $\sigma_{\rm grp}$. We insert $M_{\rm vir}$ into Eq.~(\ref{eq:c}) to estimate $c_{\rm vir}$; then use $\sigma_{\rm grp}$, $R_{\rm vir}$, and $c_{\rm vir}$ in Eq.~(\ref{eq:r0rssbar}) to estimate $\rho_0$. Using the quoted group redshift to calculate the relevant value of $\Sigma_{\rm c}$, Eqs.~(\ref{eq:ktildedef}) and~(\ref{eq:kappanfw}) allow us to obtain $\Delta\kappa$. For each system, we estimate the rough magnitude of the flexion by the root-mean-square $\sqrt{\sum\limits_{i}\left(\Delta\kappa_i/h_i\right)^2}$, including all LOS groups $i$, and estimate the corresponding deflection angle near the primary lens by $\Delta\theta\sim\sqrt{\sum\limits_{i}\left(\Delta\kappa_i/h_i\right)^2}\theta_{\rm E}^2$, taking $\theta_{\rm E}=1''$ for definiteness. 

Of the 26 systems considered in~\cite{Wilson:2017apg}, PG1115+080 does indeed come up with the highest flexion estimate of $\Delta\theta\sim0.02''$. The next highest system is HE0435, with $\Delta\theta\sim0.005''$. Altogether, 3 systems (7 systems) out of the 26 have $\Delta\theta\gtrsim0.0025''$ ($\Delta\theta\gtrsim1$~mas). 

Numerical experiments with mock analyses mimicking the main features of the pipeline of~\cite{Millon:2019slk} suggest that flexion terms are quantitatively important in the fit if the flexion-induced deflection angle is larger than $\sim1$~mas. Very roughly, we can expect that flexion degeneracy should be a concern in the same parametric regime. It is interesting to note that all of the 4 systems that take part in both of the cosmography~\cite{Millon:2019slk} and kinematics~\cite{Wilson:2017apg} campaigns, turn out to exhibit $\Delta\theta\gtrsim1$~mas.

\section{Full corner plots.}\label{app:1}
In this Appendix we collect some detailed results from the MCMC analysis. 

\begin{figure*}
	\centering
	\includegraphics[width=0.45\textwidth]{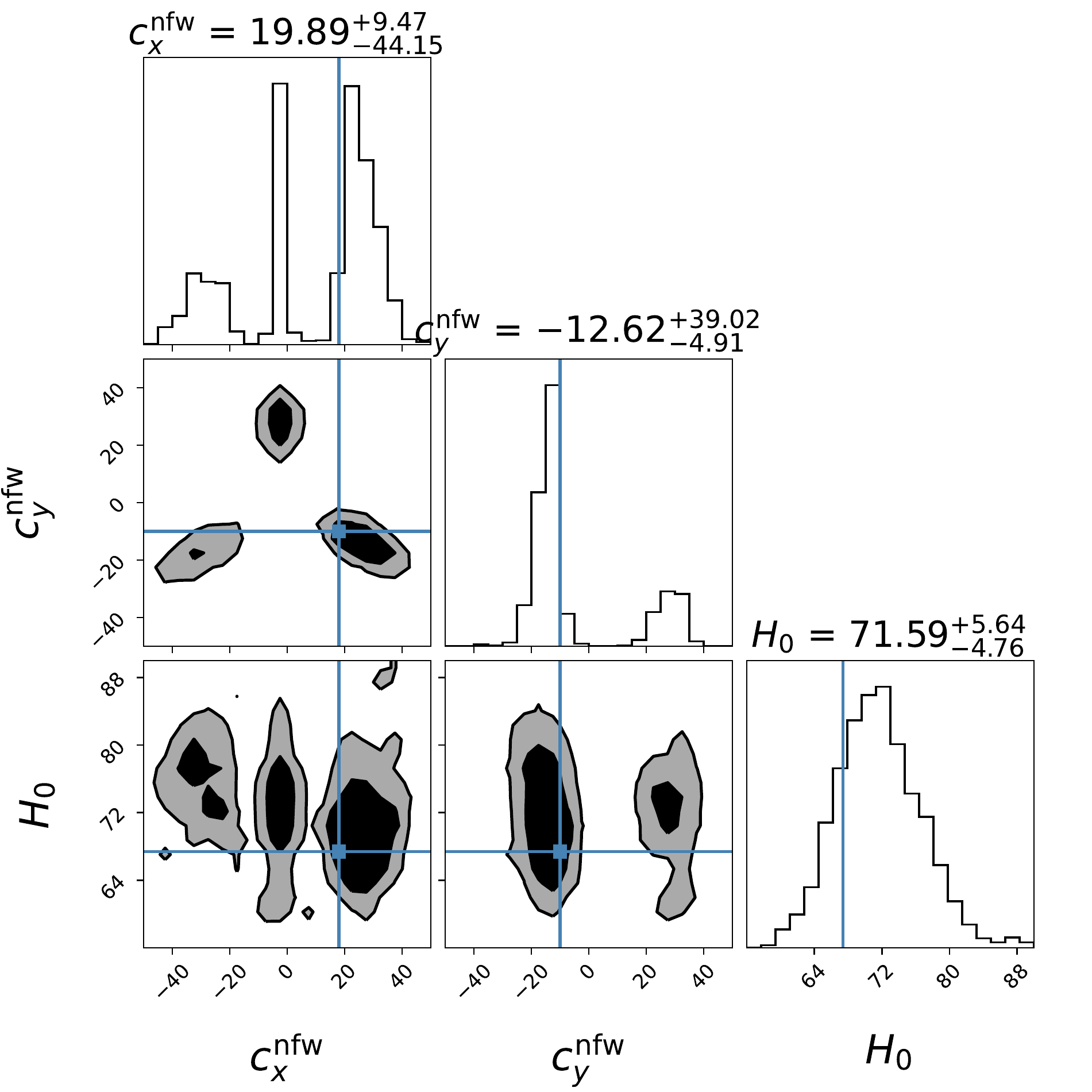}
	\includegraphics[width=0.45\textwidth]{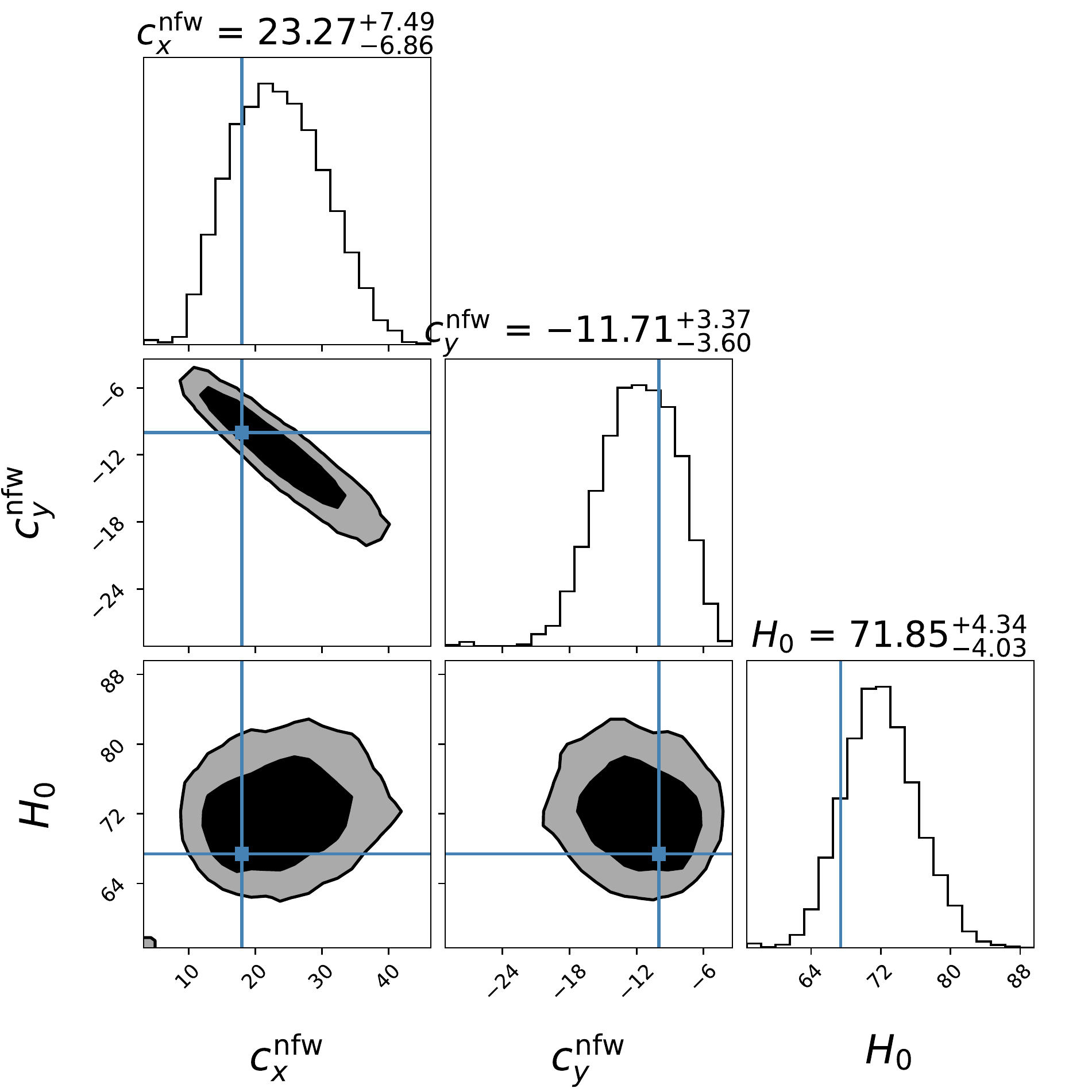}
	\includegraphics[width=0.45\textwidth]{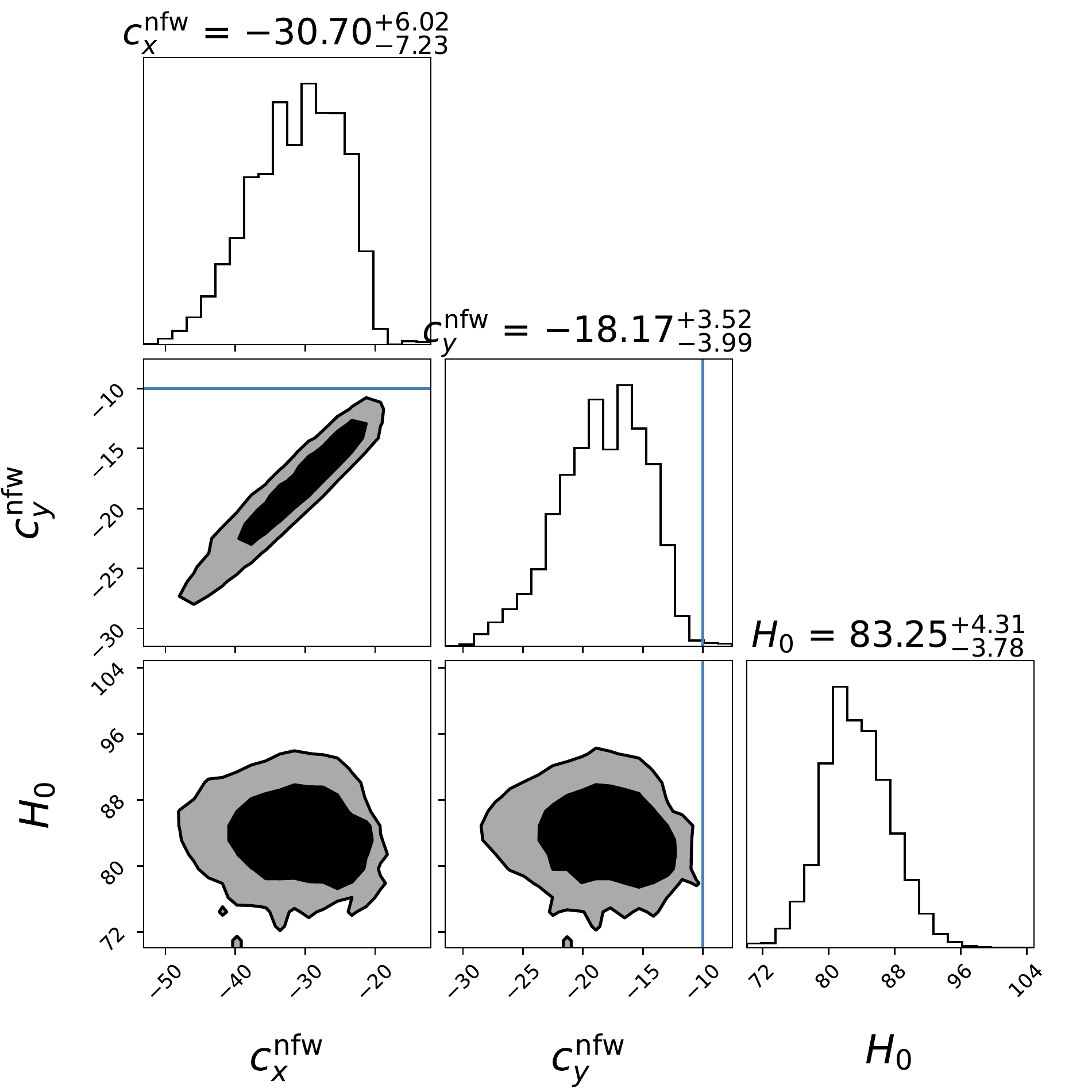}
	\includegraphics[width=0.45\textwidth]{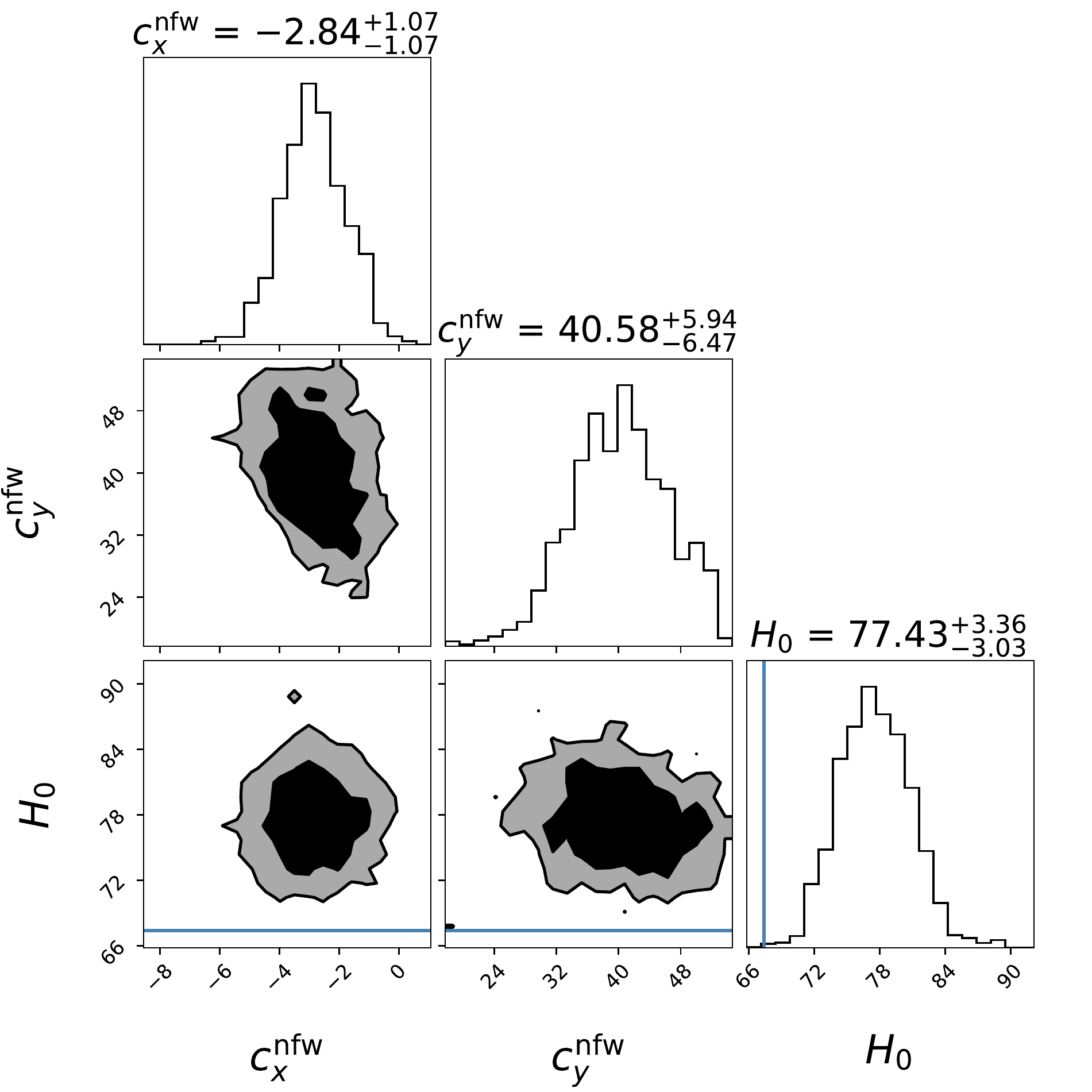}
	\caption{Mock analysis: \texttt{zeus} run (upper-right) and  \texttt{emcee} runs, all with a kinematics prior but no $ c_{\rm vir} $ prior. We note that in false minima fits, the inferred $ \theta_{\rm s} $  (not shown here) is driven to very low values \href{https://github.com/lucateo/Group_Halo_Degeneracy/tree/main}{\faGithub}.}
	\label{fig:mins_nocprior}
\end{figure*}

\begin{figure*}
	\centering
	\includegraphics[width=\textwidth]{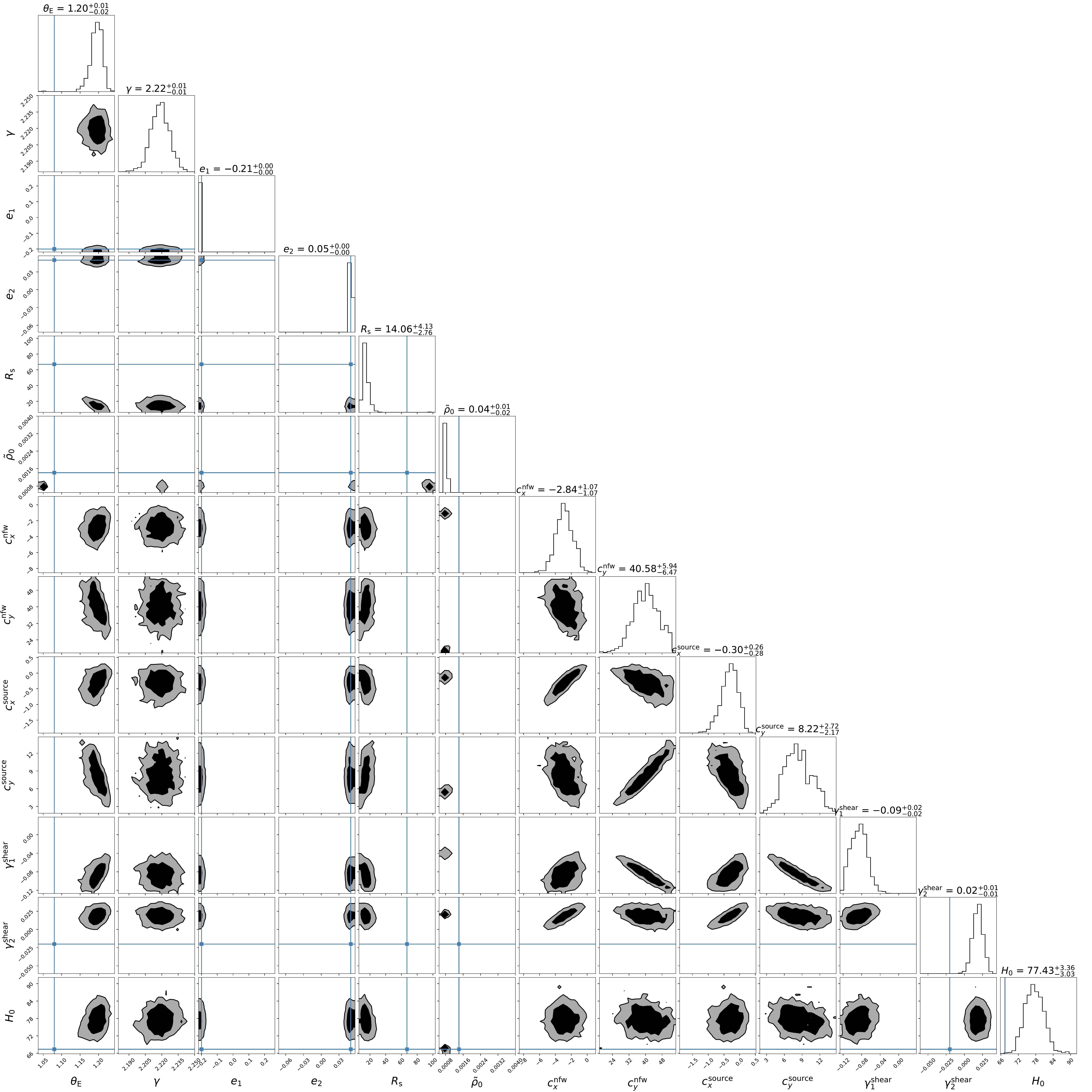}
	\caption{\texttt{emcee} run, multi-parameter view, corresponding to the run in the bottom-left panel of Fig.~\ref{fig:mins_nocprior}. The absence of a $ c_{\rm vir} $ prior causes the inference to pull towards low values of $ R_s $ near a displaced minimum. The parameter $ \tilde{\rho}_0 $ is defined as $ \tilde{\kappa}/\theta_{\rm s} $ (which is independent on $ R_{\rm s} $) \href{https://github.com/lucateo/Group_Halo_Degeneracy/blob/main/mock.ipynb}{\faGithub}.}
	\label{fig:nocprior_false_minimum_full}
\end{figure*}

In Fig.~\ref{fig:mins_nocprior} we show triangle plots in which the cosmological prior on $c_{\rm vir}$ is not included. We do this exercise in order to investigate the impact of this prior on the results. The main point to notice is that omitting the $c_{\rm vir}$ prior, the bias on $H_0$ becomes somewhat more pronounced (compare Fig.~\ref{fig:mock_inference}, that includes this prior). 
At the same time, without this prior, the best fit result for $R_{\rm s}$ in displaced (false) posterior likelihood minima is driven to small values. This point is shown by a comprehensive triangle plot in Fig.~\ref{fig:nocprior_false_minimum_full}.

\begin{figure*}
	\centering
	\includegraphics[width=0.45\textwidth]{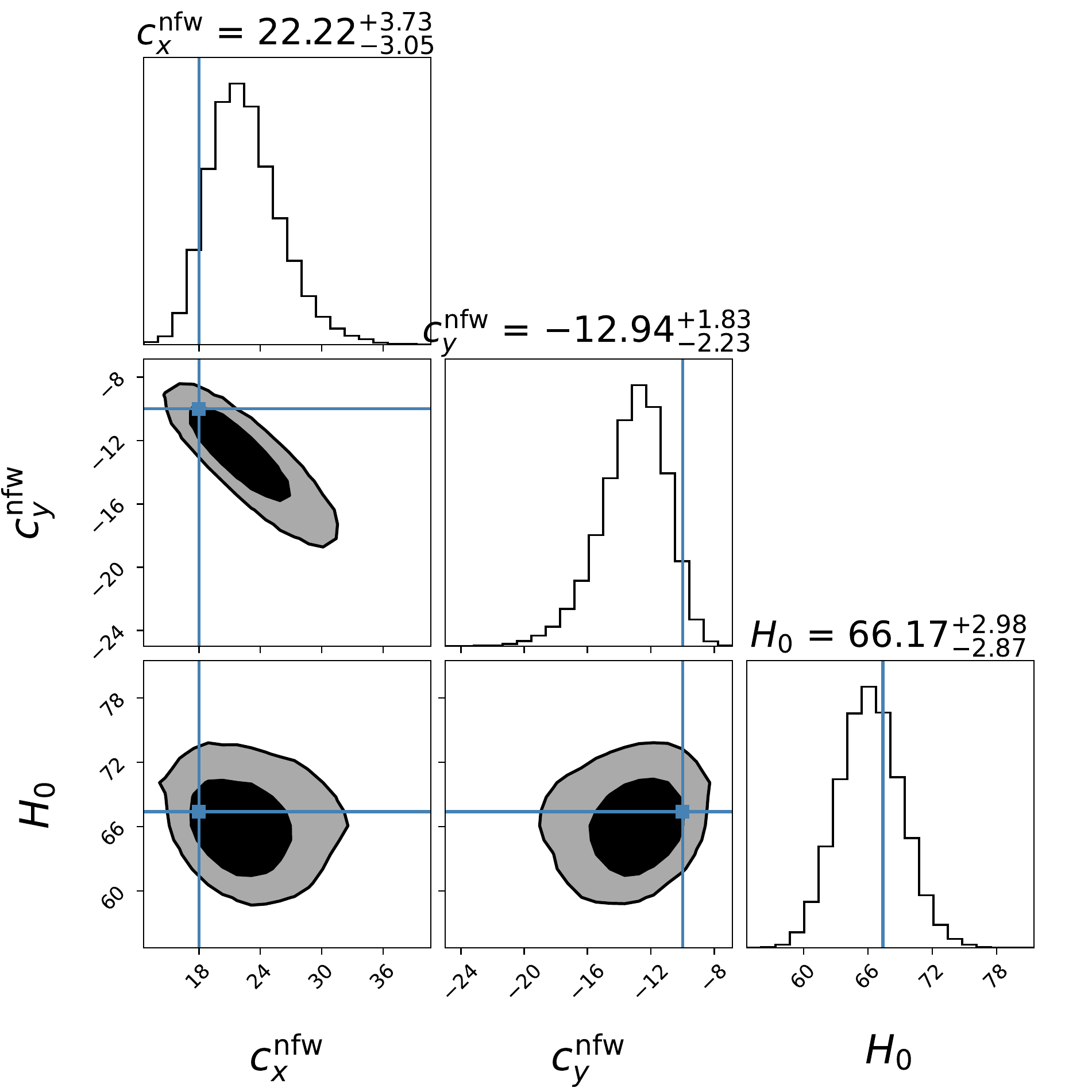}
	\includegraphics[width=0.45\textwidth]{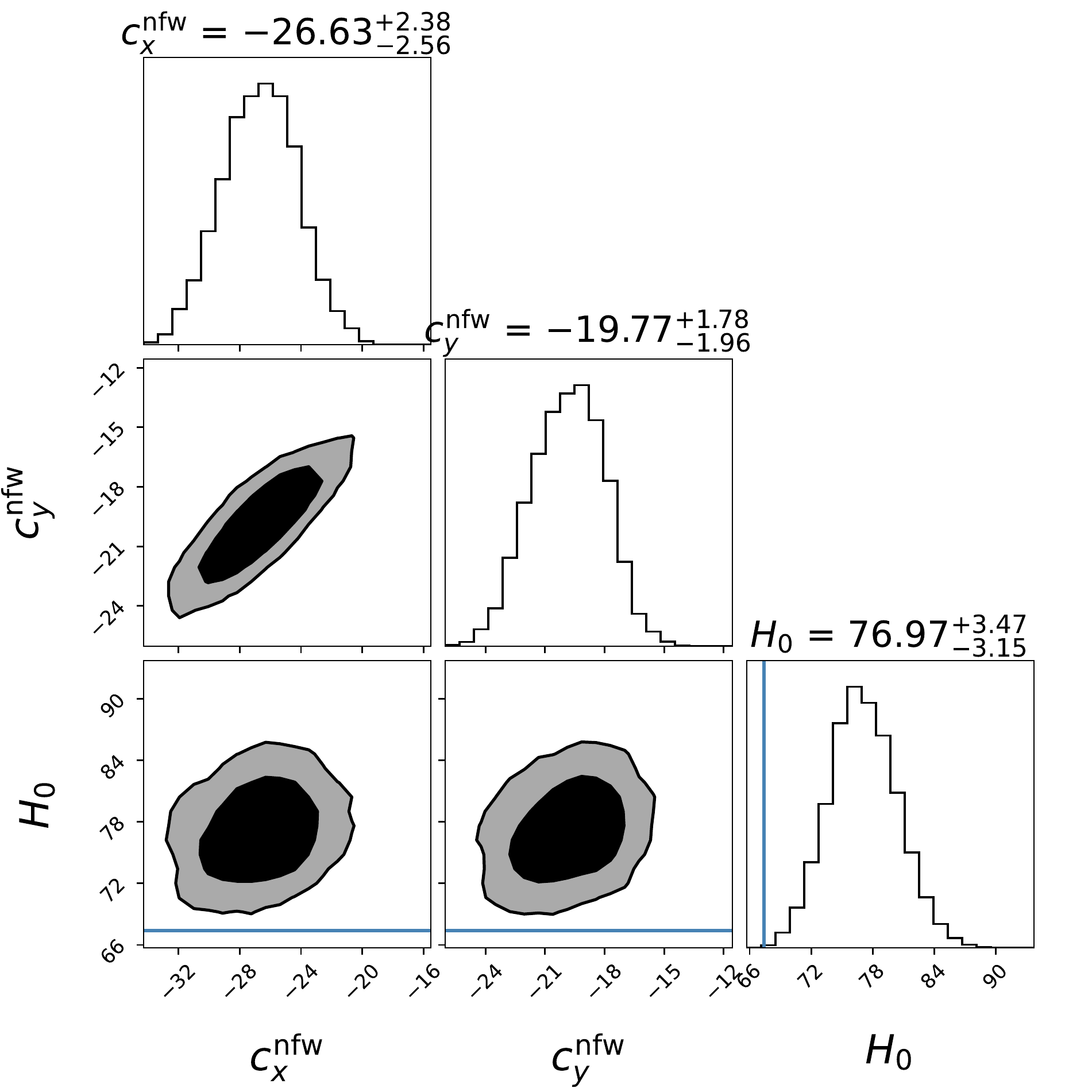}
	\includegraphics[width=0.45\textwidth]{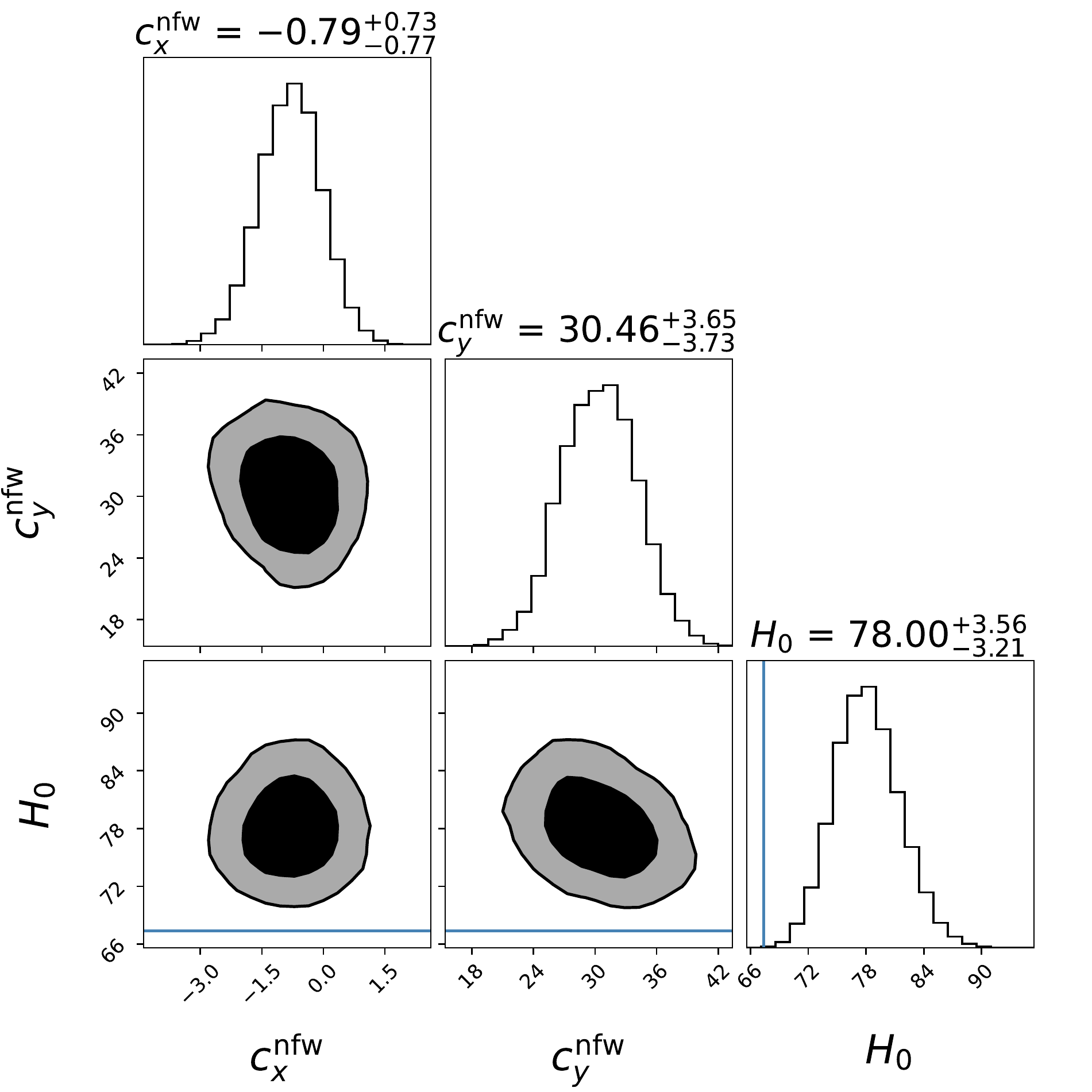}
	\caption{Mock analysis: \texttt{emcee} runs using a standard deviation of $60$~km/s on $\sigma_{\rm los}$. These results can be compared with Fig.~\ref{fig:mock_inference} from the main text, in which the uncertainty in $\sigma_{\rm los}$ was doubled \href{https://github.com/lucateo/Group_Halo_Degeneracy/blob/main/mock.ipynb}{\faGithub}.}
	\label{fig:mins_lowsigma}
\end{figure*}

In Fig.~\ref{fig:mins_lowsigma} we show triangle plots in which the kinematics prior  is enforced with a standard deviation of $ 60 $~km/s on $\sigma_{\rm los}$. These results can be compared to Fig.~\ref{fig:mock_inference} from the main text, where, as noted in Sec.~\ref{ss:extprior}, the standard deviation on $\sigma_{\rm los}$ was taken as $120$~km/s. We do not find a significant difference.

\begin{figure*}
    \centering
    \includegraphics[width=\textwidth]{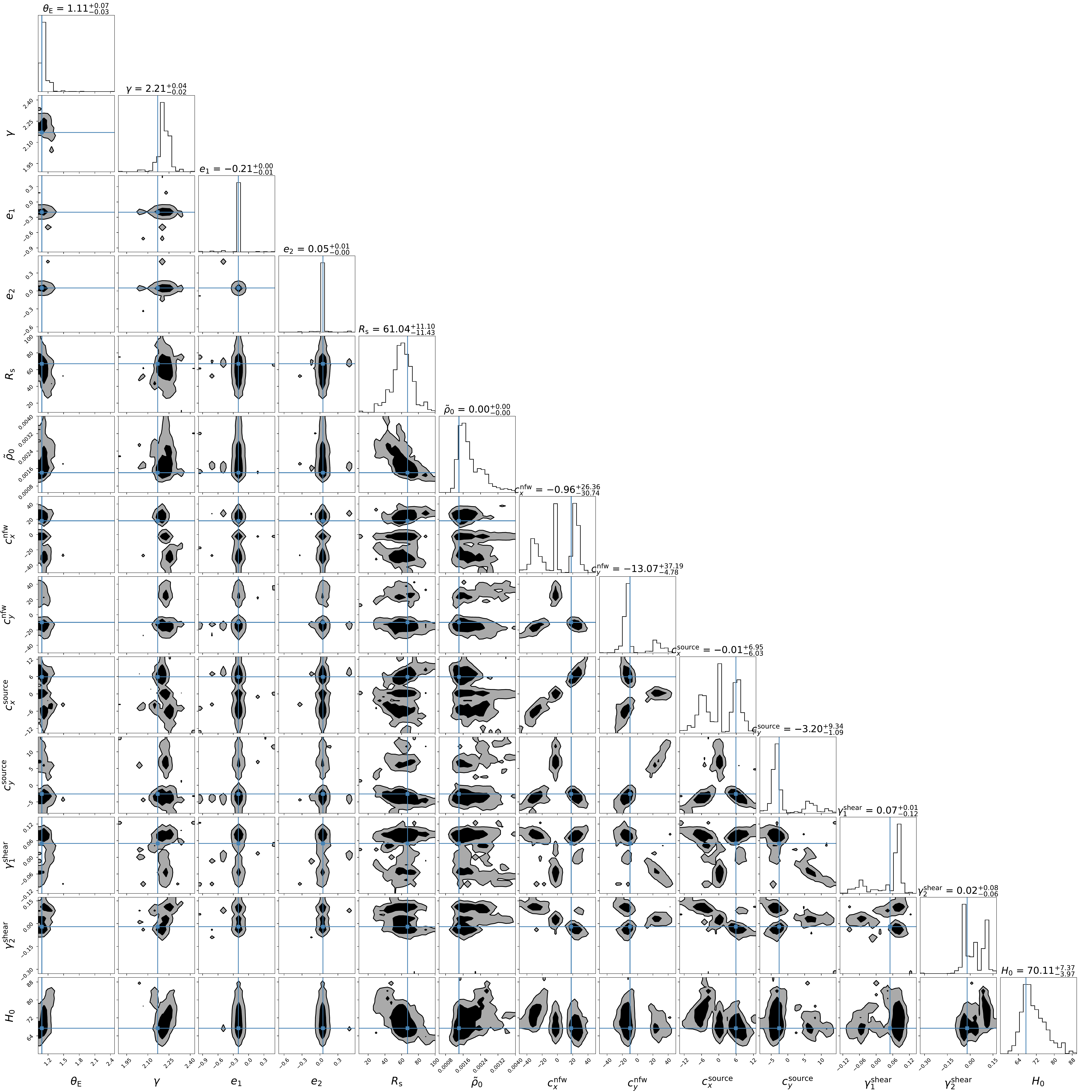}
    \caption{Mock analysis: \texttt{zeus} run, comprehensive view \href{https://github.com/lucateo/Group_Halo_Degeneracy/blob/main/mock_zeus.ipynb}{\faGithub}. A cautionary comment is needed: the run has not converged. The main point of this run is to qualitatively show the three minima structure. No quantitative results are obtained from this run.}
    \label{fig:zeusfull}
\end{figure*}

Fig.~\ref{fig:zeusfull} gives a more complete perspective on the degeneracies and the global structure of the likelihood as exposed by a \texttt{zeus} run.

\end{appendix}

\vspace{6 pt}

\bibliography{ref}

\end{document}